\documentclass[aps,prx,showpacs,english,twocolumn]{revtex4}

\usepackage[pdftex]{graphicx}
\usepackage{babel,amsmath,amssymb,float}

\usepackage{mathbbol,bbm}
\usepackage{bm,dsfont}

\newcommand{\qed}{\hspace*{\fill}$\square$}

 \newtheorem{thm}{Theorem}

\newcommand{\be}{\begin{equation}}
\newcommand{\ee}{\end{equation}}

 \newcommand{\traza}{\mathrm{Tr}}

 \newcommand{\ket}[1]{|#1\rangle}
 \newcommand{\bra}[1]{\langle #1|}

\begin{document}

\title[Short Title]{Density Matrix Topological Insulators}

\author{A. Rivas, O. Viyuela and M. A. Martin-Delgado}
\affiliation{Departamento de F\'isica Te\'orica I, Universidad Complutense, 28040 Madrid, Spain.}

\begin{abstract}
Thermal noise can destroy topological insulators (TI). However we demonstrate how TIs can be made stable in dissipative systems. To that aim, we introduce the notion of band Liouvillian as the dissipative counterpart of band Hamiltonian, and show a method to evaluate the topological order of its steady state. This is based on a generalization of the Chern number valid for general mixed states (referred to as density matrix Chern value), which witnesses topological order in a system coupled to external noise.  Additionally, we study its relation with the electrical conductivity at finite temperature, which is not a topological property. Nonetheless, the density matrix Chern value represents the part of the conductivity which is topological due to the presence of quantum mixed edge states at finite temperature. To make our formalism concrete, we apply these concepts to the two-dimensional Haldane model in the presence of thermal dissipation, but our results hold for arbitrary dimensions and density matrices.
\end{abstract}


\maketitle

\section{Introduction}
Topological insulators have emerged as a new kind of quantum phase of matter \cite{RMP1,PhysTod,RMP2,Moore}, which was predicted theoretically to exist and has been discovered experimentally \cite{TI_exp1,TI_exp3D,TI_exp2}. However, the behavior of topological insulators (TIs) subjected to dissipative dynamics has been barely explored. This is inescapable to address questions such as their robustness to thermal noise, which is crucial in assessing the feasibility of these proposals in quantum computation, spintronics, etc.

In a recent work \cite{Creutz_Ladder2012}, we have shown that certain one-dimensional
topological insulators (TI) lose the topological protection of their edge states when they are coupled to bosonic
thermal baths. This is so even when the bath interaction preserves the symmetry that protects
the existence of edge states. As a consequence, these edge states decay in time into bulk states
of a normal insulator. Thus, a very fundamental question arises: Is it possible to have stable topological
insulating states in the presence of a thermal bath? The purpose of this work is to explore this possibility by extending the concept of TI to dissipative systems. Since for dissipative systems, quantum states are generally mixed and characterized by a density matrix operator $\rho$, we shall refer to these as {\it density-matrix TIs}.

For usual TIs, the TKNN invariant \cite{TKNN} provides a characterization of fermionic time-reversal-broken (TRB) topological order in two spatial dimensions. This is done in such a way that the transverse conductivity is written in terms of a topological invariant, the Chern number, which may be related to an adiabatic change of the Hamiltonian in momentum space \cite{Hatsugai}. However, the extension of this invariant to density matrices is not straightforward. Actually, the problem of generalizing geometric concepts as distances or geometric phases to generally mixed states is highly non-trivial \cite{Ericsson,Filipp,Sjoqvist,Singh,Uhlmann}. We address this problem and construct an observable that detects the symmetry-protected topological order of a TI even if it is not in a pure, but in a general quantum mixed state. Moreover, when this general quantum state is of the form of a Gibbs state, we study the relation between this topological observable and the conductivity, and show that it reduces to the usual notion of TI in the limit of low temperature. However, we stress that the notion of a density-matrix TI is far more general, as we shall see.

\begin{figure}[t]
	 \includegraphics[width=0.45\textwidth]{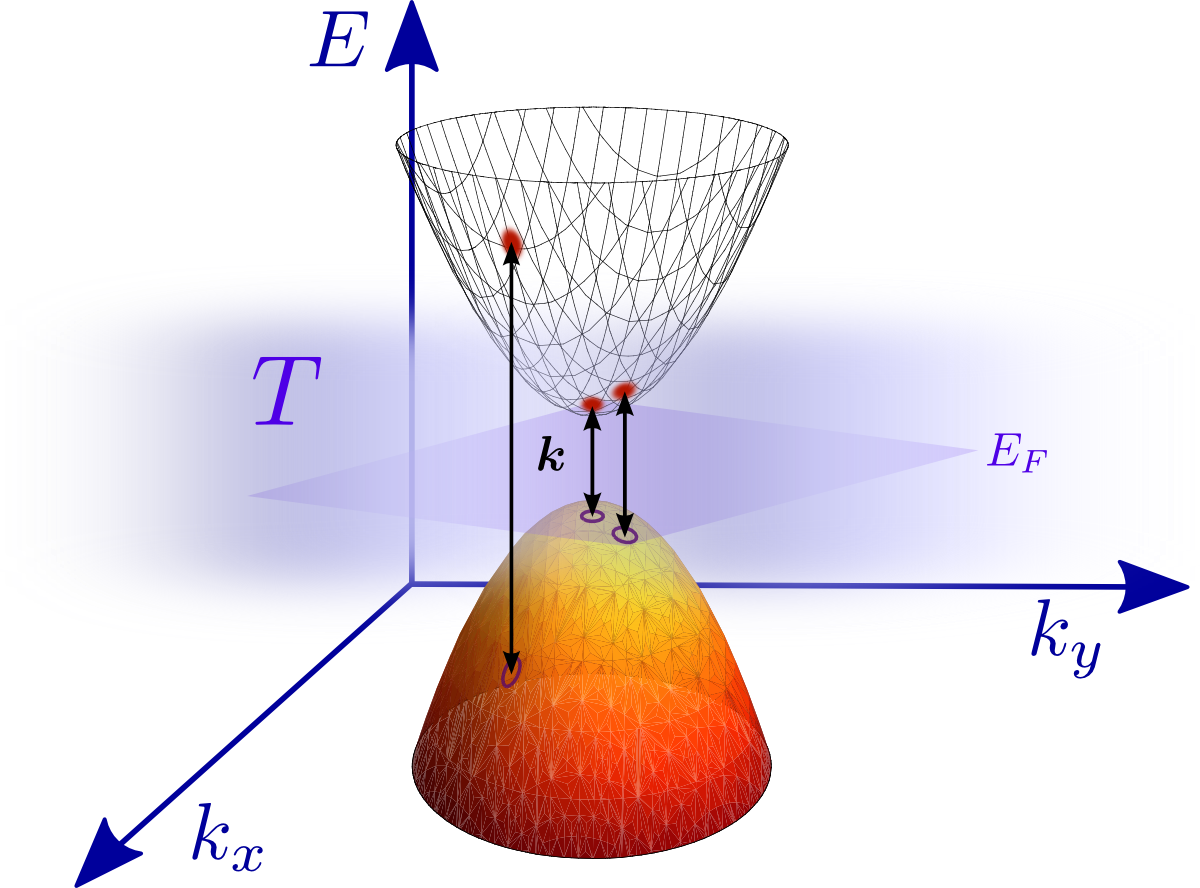}
	\caption{Pictorial image of the action of a band Liouvillian $\mathcal{L}=\sum_{\bm{k}}\mathcal{L}_{\bm{k}}$. The vertical lines denote the only possible processes involving the (initially empty) conduction band and (initially filled) valence bands, i.e. those where the momentum $\bm{k}$ is preserved. The violet fog represents some bath at a certain temperature $T$ which mediates such a processes [see Eq. \eqref{Hint}] and the plane indicates the Fermi energy $E_F$.}
	\label{fig:band_liouvillian}
\end{figure}

The paper is organized as follows. In section \ref{sec:bandLiouvillian}, we introduce the concept of band Liouvillian, which is an appropriate structure for dissipative dynamics in order to preserve topological order. Section \ref{sec:chernvalue} is devoted to construct a topological indicator for density matrices, which plays the same role as the TKNN invariant for pure states. In Section \ref{sec:LiouvillianHaldanemodel}, we analyze an example of band Liouvillian dynamics for the Haldane model of 2D TI and, subsequently, in section \ref{sec:chernvalueHaldanemodel}, we study its topological properties by using the indicator introduced in section \ref{sec:chernvalue}. Section \ref{sec:dissipativeedgestates} focuses on the behavior of this model under open boundary conditions; this leads to the appearance of mixed edge states in analogy to usual (pure) edge states of a TI in the absence of dissipation. Finally, in section \ref{sec:conductivity}, we explain the relation between  dissipative topological order and quantum Hall conductivity. Section \ref{sec:conclusions} is devoted to conclusions. In addition, technical details concerning the diagonalization of the Haldane model, derivation of master equations, and its stationary properties are left to Appendixes \ref{app_B}, \ref{app_C} and \ref{app_D}, respectively.

\section{Band Liouvillian Dynamics}
\label{sec:bandLiouvillian}

The physical problem is defined as follows. Let $H_{\rm s}$ be the system Hamiltonian representing a certain
TI. This could be constructed in an arbitrary spatial dimension, but we shall restrict in what follows
to the class of TRB insulators in two spatial dimensions. Furthermore, the TI will be subjected to the action of dissipative effects due to a thermal bath represented by a Hamiltonian $H_{\rm b}$. This bath could be general enough so as to
comprise fermionic or bosonic degrees of freedom and we assume it is initially in a thermal or Gibbs state at a certain
temperature $T$. The system-bath interaction is described by the Hamiltonian $H_{\rm s-b}$.

We consider that the state $\rho_{\rm s}$ of the TI undergoes a time evolution satisfying some Lindblad dynamical equation \cite{Alicki,GZ,BrPe,Libro} (unless otherwise stated, natural units $\hbar=k_B=1$ are taken throughout the paper):
\begin{equation}
\frac{{\rm d}\rho_{\rm s}}{{\rm d}t} = {\cal L} (\rho_{\rm s}) = -{\rm i} [H_{\rm s},\rho_{\rm s}] + {\cal D}(\rho_{\rm s}),
\label{Lindblad}
\end{equation}
where $\mathcal{L}$ is the so-called {\it Liouvillian} operator, which is composed by a first term representing the Hamiltonian evolution in the absence of system-bath interaction and a second term, the dissipator ${\cal D}$, accounting for the effect of the bath dissipation. Concretely, we shall assume that $\mathcal{L}$ is of the Davies type, obtained under the assumption of weak system-bath coupling \cite{Davies}.

We are interested in searching for sufficient conditions that the Liouvillian dynamics \eqref{Lindblad} must satisfy in order to preserve the TI phase. In the absence of dissipation, we know that a key ingredient is that the TI Hamiltonian $H_{\rm s}$ is a band Hamiltonian that  satisfies the Bloch theorem and can be decomposed as $H_{\rm s}=\sum_{\bm k\in {\rm B.Z.}} H_{\rm s}(\bm{k})$ where $\bm{k}$ denotes a crystalline momentum. Thus, it is natural to restrict our attention to Liouvillian evolutions satisfying a similar condition, $\mathcal{L}=\sum_{\bm{k}\in {\rm B.Z.}}\mathcal{L}_{\bm{k}}$, where each $\mathcal{L}_{\bm{k}}$ only involves fermionic operators with crystalline momentum $\bm{k}$; we shall refer to these as {\it band Liouvillians}. Basically, a Liouvillian of this kind describes processes in such a way that the momenta of the fermions are not changed (up to a vector $\bold{G}$ of the reciprocal lattice), a pictorial image is sketched in Fig. \ref{fig:band_liouvillian}. As a consequence, they present invariance under space translations and every $\mathcal{L}_{\bm{k}}$ satisfies
\begin{equation}
T(\bm{a})\mathcal{L}_{\bm{k}}(\rho) T^\dagger(\bm{a})=\mathcal{L}_{\bm{k}}\left[T(\bm{a})\rho T^\dagger(\bm{a})\right],
\label{tinvariance}
\end{equation}
where $T(\bm{a})={\rm e}^{-{\rm i} \bm{a}\cdot\hat{\bm {k}}}$ is the operator that translates a point with coordinate $\bm{r}$ to the point $\bm{r}+\bm{a}$ on the lattice.

An analogy to the Bloch theorem for this kind of Liovillians characterizes its steady states.
\begin{thm}
Consider a band Liouvillian $\mathcal{L}=\sum_{\bm{k}\in {\rm B.Z.}}\mathcal{L}_{\bm{k}}$. If each $\mathcal{L}_{\bm{k}}$ has a unique stationary state, it has the form
\begin{equation}
\rho_{\rm ss}=\lambda_0\ket{0}\bra{0}+\sum_{\bm{k},\alpha,\beta}\lambda^{\bm{k}}_{\alpha\beta}\ket{1_{\alpha,\bm{k}}}\bra{1_{\beta,\bm{k}}}.
\label{rhoss1}
\end{equation}
Here, $\alpha$ and $\beta$ denote additional quantum numbers (band indexes, spin indexes, lattice indexes, etc.), and $\ket{1_{\alpha,\bm{k}}}\equiv\ket{u_{\alpha,\bm{k}}}$ denotes a particle in the Bloch state with momentum $\bm{k}$ and additional quantum number $\alpha$, $T(\bm{r})\ket{u_{\alpha,\bm{k}}}={\rm e}^{-{\rm i} \bm{r}\cdot\bm {k}}\ket{u_{\alpha,\bm{k}}}$.
\end{thm}

\noindent {\it Proof:}
Consider $\rho_{\rm ss}$ to be the steady state of some $\mathcal{L}_{\bm k}$:
\begin{equation}
\mathcal{L}_{\bm k}(\rho_{\rm ss})=0.
\end{equation}
By applying the translation operator on both sides and using \eqref{tinvariance}, we obtain
\begin{equation}
\mathcal{L}_{\bm k}\left[T(\bm{a})\rho_{\rm ss} T^\dagger(\bm{a})\right]=0.
\end{equation}
Thus, $\rho_{\rm ss}':=T(\bm{a})\rho_{\rm ss} T^\dagger(\bm{a})$ is also a steady state of the system. Since by assumption $\rho_{\rm ss}$ is unique, $\rho_{\rm ss}'=\rho_{\rm ss}$, so that $[T(\bm{a}),\rho_{\rm ss}]=0$ and $T(\bm{a})$ and $\rho_{\rm ss}$ share the same set of eigenvectors.
\hfill $\square$

It is worth noticing that as a difference with the case of  pure states, the translational symmetry of a Liovillian does not necessarily implies steady states with well-defined crystalline momentum $\bm{k}$. They can be a convex mixture of states with different well-defined momenta \eqref{rhoss1}. However, since the subspace with well-defined momentum $\bm{k}$ is invariant under the action of $\mathcal{L}_{\bm k}$, if the initial state has a well-defined momentum $\bm{k}$ (for instance a particle with well-defined momentum in one of the bands of the Hamiltonian), then the steady state under $\mathcal{L}_{\bm k}$ will have well-defined momentum as well,
\begin{equation} \label{rhoss2}
\rho_{\rm ss}^{\bm{k}}=\lambda_0\ket{0}\bra{0}+ \sum_{\alpha,\beta}\lambda_{\alpha\beta}^{\bm{k}}\ket{1_{\alpha,\bm{k}}}\bra{1_{\beta,\bm{k}}}.
\end{equation}
Thus, the steady state $\rho_{\rm ss}$ of the total Liouvillian $\mathcal{L}=\sum_{\bm{k}\in {\rm B.Z.}}\mathcal{L}_{\bm{k}}$ will be of the form
\begin{equation}
\rho_{\rm ss}=\bigotimes_{\bm k} \rho_{\rm ss}^{\bm{k}}, \quad \rho_{\rm ss}^{\bm{k}}:=\lambda_0\ket{0}\bra{0}+ \sum_{\alpha,\beta}\lambda_{\alpha\beta}^{\bm{k}}\ket{1_{\alpha,\bm{k}}}\bra{1_{\beta,\bm{k}}}.
\label{Bloch_Lindbland}
\end{equation}
The coefficients $\lambda_0$ and $\lambda_{\alpha\beta}^{\bm{k}}$ depend on the particular steady state as a result of the dissipative dynamics \eqref{Lindblad}. For instance, if the steady state turns out to be a thermal state density matrix, then they will be given by Gibbs weights \eqref{steady}.

\section{Chern Connections for Density Matrices}
\label{sec:chernvalue}

In order to construct a topological indicator for the generally mixed state $\rho_{\rm ss}$, we cannot use the usual Berry connection as in the formulation of the Chern number, because it is defined just for pure states. Regarding density matrices, there is not a unique natural extension of the Berry connection and the Berry phase \cite{Ericsson,Filipp,Sjoqvist,Singh,Uhlmann}. However, the band Liouvillian structure allows for the construction of a Berry-type connection $A_i^{\rho}$ for the density-matrix steady states, in such a way that the integral of its curvature form, $F_{ij}^{\rho}$,  gives a topological indicator which we refer to as {\it density-matrix Chern value}. To construct such an indicator, we use purifications, which is a method that allows us to extend quantities defined for pure states to general mixed states.

Generally speaking, for a density matrix $\rho$ acting in a Hilbert space $\mathcal{H}$, a purification $\ket{\Phi^\rho}$ is a pure state in an extended Hilbert space $\ket{\Phi^\rho}\in\mathcal{H}_A\otimes\mathcal{H}$ such that
\begin{equation}\label{puri}
\rho=\traza_A\left(\ket{\Phi^\rho}\bra{\Phi^\rho}\right).
\end{equation}
In other words, mixed states can always be seen as pure states of a larger system such that we only have access to partial information of it.

Given some $\rho$ there are infinitely many states $\ket{\Phi^\rho}$ which fulfill \eqref{puri}. Without loss of mathematical generality, we take the ancillary space $\mathcal{H}_A$ to have the same dimension $d$ as $\mathcal{H}$ for the system \cite{footnote1}, then any purification $\ket{\Phi^\rho}$ can be written as
\begin{equation}\label{puridesc}
\ket{\Phi^\rho}=(U_A\otimes\tilde{\rho})\ket{\Omega},
\end{equation}
where $U_A$ is a unitary operator, $\tilde{\rho}\tilde{\rho}^\dagger=\rho$ and
\begin{equation}
\ket{\Omega}:=\sum_{\alpha=1}^d\ket{v_\alpha}\otimes\ket{v_\alpha},
\end{equation}
is a (unnormalized) maximally entangled state, with $\{\ket{v_j}\}$ an orthonormal basis of $\mathcal{H}$. From the Schmidt decomposition of $\ket{\Phi^\rho}$ it follows that \eqref{puridesc} is the most general form for a purification of $\rho$ \cite{NC00}.

By using the spectral decomposition of $\rho=\sum_\alpha p_\alpha\ket{\psi_\alpha}\bra{\psi_\alpha}$, we may write $\tilde{\rho}$ as
\begin{equation} \label{tilderho}
\tilde{\rho}=\sum_\alpha\sqrt{p_\alpha}\ket{\psi_\alpha}\bra{\varphi_\alpha},
\end{equation}
where $\{\ket{\varphi_\alpha}\}$ is also an orthonormal basis which is considered to be arbitrary. Therefore, given some $\rho$ there is a freedom for the choice of $U_A$ and the basis $\{\ket{\varphi_\alpha}\}$ for its purification $\ket{\Phi^\rho}$.
 
The next theorem is particularly important in the construction of a Berry-type connection for density matrices.


\begin{thm}
Consider the steady state of a band Liouvillian, Eq. \eqref{Bloch_Lindbland}, we define a Berry-type connection for $\rho_{\rm ss}^{\bm{k}}$ through one of its purifications $\ket{\Phi^\rho_{\bm{k}}}$ as
\begin{equation}\label{Arho}
A^\rho_i(\bm{k}):={\rm i}\langle \Phi^\rho_{\bm{k}}|\partial_i\Phi^\rho_{\bm{k}}\rangle, \quad \text{for\ \ }\rho_{\rm ss}^{\bm{k}}=\traza_A\left(\ket{\Phi^\rho_{\bm{k}}}\bra{\Phi^\rho_{\bm{k}}}\right).
\end{equation}
Here, the notation is $\partial_i:=\partial_{k_i}$.
Under the assumption that $U_A$ and $\{\ket{\varphi_i}\}$ are independent of momentum  $\bm{k}$  of the steady state,
the connection \eqref{Arho} is unique and does not depend on the purification. Explicitly, it takes the following form in terms of the spectral decomposition of $\rho_{\rm ss}^{\bm{k}}=\sum_{\alpha}p_\alpha^{\bm{k}}\ket{\psi_{\alpha,\bm{k}}}\bra{\psi_{\alpha,\bm{k}}}$:
\begin{equation}\label{ArhoSimplified2}
A^\rho_i(\bm{k})={\rm i}\sum_{\alpha}p_\alpha^{\bm{k}}\langle\psi_{\alpha,\bm{k}}|\partial_i\psi_{\alpha,\bm{k}}\rangle.
\end{equation}
\end{thm}

\noindent {\it  Proof:}  Indeed, the general form \eqref{puridesc} for a purification of $\rho_{\rm ss}^{\bm{k}}$ reads as
\begin{equation}
\ket{\Phi^\rho_{\bm{k}}}=\sum_\alpha\sqrt{p_\alpha^{\bm{k}}}\left(U_A\otimes\ket{\psi_{\alpha,\bm{k}}}\bra{\varphi_\alpha}\right)\ket{\Omega},
\label{Purification}
\end{equation}
where we have used \eqref{tilderho}.
Taking the derivative $\partial_i:=\partial_{k_i}$ in \eqref{Purification} and computing the overlap
\begin{equation}
\begin{split}
\langle \Phi^\rho_{\bm{k}}|\partial_i\Phi^\rho_{\bm{k}}\rangle &=  \sum_{\alpha,\beta}\sqrt{p_\beta^{\bm{k}}}\bra{\Omega} \\
&\left[ \left(\partial_i\sqrt{p_\alpha^{\bm{k}}}\right)\left(\mathbb{1}\otimes\ket{\varphi_\beta}\langle\psi_{\beta,\bm{k}}|\psi_{\alpha,\bm{k}}\rangle\bra{\varphi_\alpha}\right) \right.\\
& \left.+\sqrt{p_\alpha^{\bm{k}}}\left(\mathbb{1}\otimes\ket{\varphi_\beta}\langle\psi_{\beta,\bm{k}}|\partial_i\psi_{\alpha,\bm{k}}\rangle\bra{\varphi_\alpha}\right) \right] \ket{\Omega}.
\end{split}
\end{equation}
Since $\bra{\Omega}(\mathbb{1}\otimes A)\ket{\Omega}=\traza(A)$, we obtain
\begin{equation}\label{ArhoSimplified}
A^\rho_i(\bm{k})={\rm i}\sum_{\alpha}\sqrt{p_\alpha^{\bm{k}}}\left(\partial_i\sqrt{p_\alpha^{\bm{k}}}\right)+p_\alpha^{\bm{k}}\langle\psi_{\alpha,\bm{k}}|\partial_i\psi_{\alpha,\bm{k}}\rangle,
\end{equation}
which is independent of $U_A$ and $\{\ket{\varphi_i}\}$. Moreover, note that in the pure state case $\rho_{\rm ss}^{\bm{k}}=\ket{\psi_{\bm{k}}}\bra{\psi_{\bm{k}}}$ we recover the Berry connection $A^\rho_i(\bm{k})=A_i(\bm{k})={\rm i}\langle \psi_{\bm{k}}|\partial_{i} \psi_{\bm{k}}\rangle$ \cite{Berry}.
In addition, the first term on the right hand side of \eqref{ArhoSimplified} vanishes by taking into account that $\sum_{\alpha}p_\alpha^{\bm{k}}=1$.
Hence, \eqref{ArhoSimplified} can be simply written as \eqref{ArhoSimplified2}.
\hfill $\square$

Once this {\it purified connection} \eqref{ArhoSimplified2} is defined, we may obtain the (Abelian) curvature form through
\begin{equation}\label{Fdef}
F_{ij}^\rho(\bm{k}):=\partial_i A_j^\rho(\bm{k})-\partial_j A_i^\rho(\bm{k}),
\end{equation}
and construct a density-matrix topological indicator ${\rm n}_{\rm Ch}^\rho$ associated with the steady state $\rho_{\rm ss}^{\bm{k}}$ via the first Chern class \cite{EGH} of this connection:
\begin{equation}\label{nrho}
{\rm n}_{\rm Ch}^\rho:=\frac{1}{4\pi} \traza\left[\int_{{\rm T}^2} F^\rho_{ij}(\bm{k}) dk_i\wedge dk_j\right].
\end{equation}
It is convenient to compute the different contributions that appear in the explicit expression of \eqref{Fdef} using \eqref{ArhoSimplified2}:
\begin{equation}\label{curvature}
F_{ij}^\rho(\bm{k})=\sum_{\alpha}\left[p_\alpha^{\bm{k}}F_{ij}^{\alpha}(\bm{k})+\left(\partial_ip_\alpha^{\bm{k}}\right) A^{\alpha}_j(\bm{k})-\left(\partial_jp_\alpha^{\bm{k}}\right) A^{\alpha}_i(\bm{k})\right].
\end{equation}
Note that from this equation is not manifestly clear the U(1) gauge invariance of the curvature, but this can be proven by performing a gauge transformation and making use of the property $\sum_{\alpha}p_\alpha^{\bm{k}}=1$. In addition, if $N$ is the dimension of the steady state $\rho_{\rm ss}^{\bm k}$, the curvature is not ${\rm U}(1)^N$ gauge invariant, however, that is not the case with the Chern value \eqref{nrho}, which is fully invariant. A proof of this fact is given in Appendix \ref{app_A}.

Thus, one of the main results of this work is the construction of this object ${\rm n}_{\rm Ch}^\rho$, which characterizes the topological structure of insulators in the presence of dissipation. Furthermore, by taking into account Eq. \eqref{curvature}, ${\rm n}_{\rm Ch}^\rho$ can be written as
\begin{equation}\label{nrho2}
\begin{split}
{\rm n}_{\rm Ch}^\rho= & \frac{1}{2\pi}\int_{{\rm T}^2}F_{12}^{\rho}(\bm{k})d^2\bm{k}= \frac{1}{2\pi}\sum_{\alpha}\int_{{\rm T}^2}p_\alpha^{\bm{k}}F_{12}^{\alpha}(\bm{k})d^2\bm{k} \\
 + &\frac{1}{2\pi}\sum_{\alpha}\int_{{\rm T}^2}\left[\left(\partial_1p_\alpha^{\bm{k}}\right)A^{\alpha}_2(\bm{k})-\left(\partial_2p_\alpha^{\bm{k}}\right)A^{\alpha}_1(\bm{k})\right]d^2\bm{k},
\end{split}
\end{equation}
A non-vanishing ${\rm n}_{\rm Ch}^\rho$ witnesses a topological non-trivial order present in $\rho_{\rm ss}^{\bm{k}}$. Since for the pure case the connection \eqref{ArhoSimplified2} becomes the usual Berry connection, if the steady state is a pure Bloch state $\rho_{\rm ss}^{\bm{k}} = \ket{u_{\alpha,\bold{k}}}\bra{u_{\alpha,\bold{k}}}$, we recover the standard TKNN topological invariant (Chern number).

The density-matrix Chern value ${\rm n}_{\rm Ch}^\rho$, Eq. \eqref{nrho2}, has two different terms. The first one is a weighted integration of curvatures for different bands. This term has no topological meaning on its own and it does not distinguish between phases with or without topological order. The second term represents a correction to the value given by the first one that provides the topological character to ${\rm n}_{\rm Ch}^\rho$. In addition, both terms have a physical meaning which will be explained in section \ref{sec:conductivity}.

\begin{figure}[t]
\includegraphics[width= 0.48\textwidth]{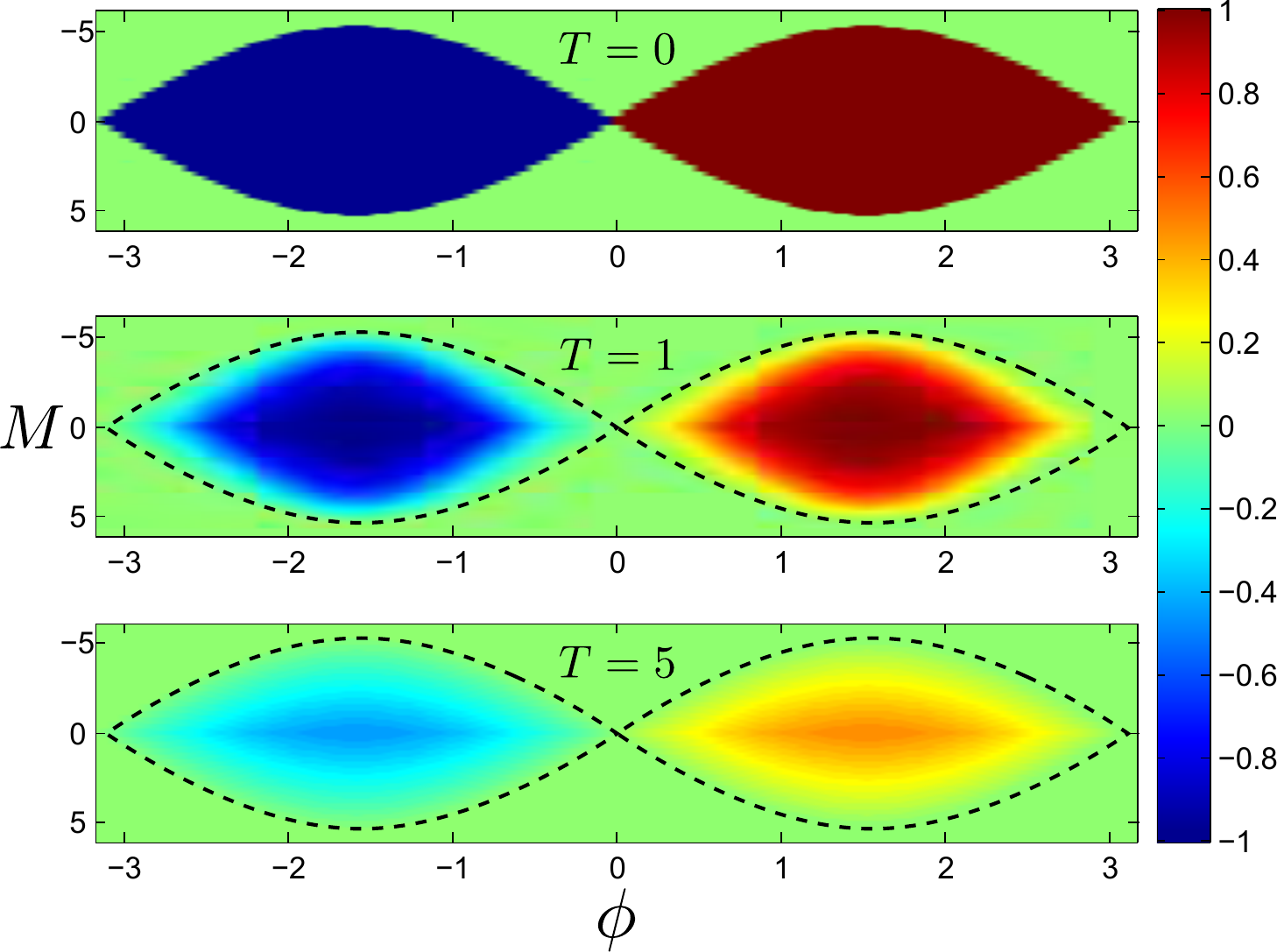}
\caption{Color map depicting the Chern value in the Haldane model with dissipation, for different values of $\phi$, $M$, and bath temperature $T$ (in units of $t_2=1$). As $T$ increases, the Chern value decreases (in absolute value), and for $T=0$ we recover the phase diagram obtained by Haldane \cite{Haldane}. The dashed black lines enclose the region displaying topological order at $T=0$, so that all nonvanishing Chern values are inside of this region for any $T$. Approximately $T=1$ and $T=5$ correspond to less than $10\%$ and $50\%$ of the gap respectively.}
\label{PhaseDiagram}
\end{figure}

The name {\it Chern value} responds to the fact that despite its topological origin, it may not be an integer for a general mixed state. The reason is very fundamental; the space of density matrices $\rho$ is a convex space, which means that a convex combination of density matrices $\rho_1$ and $\rho_2$, $\rho=p_1\rho_1 + p_2\rho_2$ is also a mixed state. Due to the Abelian character of the curvature form $F_{ij}^{\rho}$ [Eq. \eqref{Fdef}], ${\rm n}_{\rm Ch}^\rho=p_1{\rm n}_{\rm Ch}^{\rho_1}+p_2{\rm n}_{\rm Ch}^{\rho_2}$. Therefore, since the weights $p_1,p_2\in \mathds{R}$ with $p_1+p_2=1$, then ${\rm n}_{\rm Ch}^\rho \in \mathds{R}$ as well. Nevertheless, this will not be an obstacle to use the Chern value to detect topological properties of insulator states. Note that there are other quantities in the literature which also reflect topological properties but are not integers numbers, for example the Aharonov-Bohm phase, see also \cite{HaldaneFLiquid}. In forthcoming sections we will apply this formalism to the case of the Haldane model in 2D which is a prototype of TRB topological insulator \cite{Haldane}.

\section{Band Liouvillian for the Haldane Model}
\label{sec:LiouvillianHaldanemodel}

We can apply our previous formalism to the Haldane model of 2D TI. This is a graphenelike model based on a honeycomb lattice with nearest-neighbor and next-nearest-neighbor couplings. For periodic boundary conditions, the Haldane Hamiltonian in the reciprocal space is given by
\begin{equation}
H_{\rm s}=\sum_{\bm k\in {\rm B.Z.}}(a^{\dagger}_{\bm k},b^{\dagger}_{\bm k})H({\bm k})\begin{pmatrix} a_{\bm k}\\ b_{\bm k} \end{pmatrix}=\sum_{\bm k\in {\rm B.Z.}}E_1^{\bm k}c^\dagger_{\bm k}c_{\bm k}+E_2^{\bm k}d^\dagger_{\bm k}d_{\bm k}.
\label{Hs}
\end{equation}
Here, $a_{\bm k}$ and $b_{\bm k}$ correspond to the two species of fermions associated with the triangular sublattices of a honeycomb lattice, and $c_{\bm k}$ and $d_{\bm k}$ are the fermionic modes which diagonalize the Hamiltonian with eigenvalues $E_1^{\bm k}$ and $E_2^{\bm k}$, respectively. For more details, we refer to Appendix \ref{app_B}.

We shall assume a local fermionic bath model, with quadratic coupling of the form
\begin{equation} \label{Hint}
H_{\rm s-b}:=\sum_{i,\bm{r}}g^i(a^{\dagger}_{\bm{r}}\otimes A^{i}_{\bm{r}}+a_{\bm{r}}\otimes A^{i{\dagger}}_{\bm{r}}+b^{\dagger}_{\bm{r}}\otimes B^{i}_{\bm{r}}+b_{\bm{r}}\otimes B^{i{\dagger}}_{\bm{r}}),
\end{equation}
where $\bm{r}$ denotes the point in the sublattices and $A^{i}_{\bm{r}}$ and $B^{i}_{\bm{r}}$ the bath fermion operators coupled with the two species $a_{\bm r}$ and $b_{\bm r}$ respectively. This model could effectively describe situations such as; i) non-controllable tunneling of electrons between the TI and some surrounding material, ii) particle losses in simulated  topological phases with cold fermionic atoms in optical lattices, or electron transitions to high-energy levels not well-described under the tight-binding approximation.

The detailed derivation of the Liouvillian equation (master equation) in the weak coupling limit for this systems is explained in Appendix \ref{app_C}; the final result turns out to be
\begin{equation}\label{masterEq}
\begin{split}
&\frac{d\rho_{\rm s}(t)}{dt}=\sum_{\bm{k}}{\cal L}_{\bm{k}}[\rho_{\rm s}(t)] =\sum_{\bm{k}}\bigg(-{\rm i}[H_{\bm{k}},\rho_{\rm s}(t)] \\
&+\gamma(E_1^{\bm{k}})\bar{n}_F(E_1^{\bm{k}})\Big(c^{\dagger}_{\bm{k}}\rho_{\rm s}(t)c_{\bm{k}}-\frac{1}{2}\{c_{\bm{k}}c^{\dagger}_{\bm{k}},\rho_{\rm s}(t)\}\Big)+\\
&+\gamma(E_1^{\bm{k}})[1-\bar{n}_F(E_1^{\bm{k}})]\Big(c_{\bm{k}}\rho_{\rm s}(t)c^{\dagger}_{\bm{k}}-\frac{1}{2}\{c^{\dagger}_{\bm{k}}c_{\bm{k}},\rho_{\rm s}(t)\}\Big) \\
&+\gamma(E_2^{\bm{k}})\bar{n}_F(E_2 ^{\bm{k}})\Big(d^{\dagger}_{\bm{k}}\rho_{\rm s}(t)d_{\bm{k}}-\frac{1}{2}\{d_{\bm{k}}d^{\dagger}_{\bm{k}},\rho_{\rm s}(t)\}\Big)+\\
&+\gamma(E_2^{\bm{k}})[1-\bar{n}_F(E_2^{\bm{k}})]\Big(d_{\bm{k}}\rho_{\rm s}(t)d^{\dagger}_{\bm{k}}-\frac{1}{2}\{d^{\dagger}_{\bm{k}}d_{\bm{k}}, \rho_{\rm s}(t)\}\Big)\bigg).
\end{split}
\end{equation}
Here,
\begin{equation}
\bar{n}_F(E):=\frac{1}{{\rm e}^{\beta E}+1}, \quad
\gamma(\omega):=2\pi J(\omega),
\end{equation}
where $J(\omega)$ is the bath spectral density.

It is important to emphasize that this Liouvillian  \eqref{masterEq} fulfills the conditions of a band Liouvillian $\mathcal{L}=\sum_{\bm{k}\in {\rm B.Z.}}\mathcal{L}_{\bm{k}}$. Moreover, it is quadratic in fermionic operators and its unique steady state is the Gibbs state $(\beta=1/T)$
\begin{equation}
\rho_{\beta}=\frac{\text{e}^{-\beta H_{\rm s}}}{Z}=\bigotimes_{\bm{k}}\rho_{\rm ss}^{\bm{k}}=\bigotimes_{\bm{k}}\bigg(\frac{{\rm e}^{-\beta E_1^{\bm{k}}c^{\dagger}_{\bm{k}}c_{\bm{k}}}}{1+\text{e}^{-\beta E_1^{\bm{k}}}}\bigg)\bigg(\frac{{\rm e}^{-\beta E_2^{\bm{k}}d^{\dagger}_{\bm{k}}d_{\bm{k}}}}{1+\text{e}^{-\beta E_2^{\bm{k}}}}\bigg),
\label{steady}
\end{equation}
that has the form of \eqref{Bloch_Lindbland} corresponding to a band Liouvillian. Note that in the limit $T\rightarrow0$, $\rho_{\beta}$ approaches the Fermi sea where the lower band $(c_{\bm{k}})$ is fully occupied and the upper band $(d_{\bm{k}})$ is completely empty (see Appendix \ref{app_D} for more details).

\section{Chern Value of the Steady State}
\label{sec:chernvalueHaldanemodel}

We have obtained that the steady state of the Liouvillian \eqref{masterEq} is a product of states $\rho^{\bm{k}}_{\rm ss}$ with well-defined momentum. Thus, the (parallel) transport along \emph{${\bm k}$-space} of each of these states is well defined and, hence, the state characterization by a density-matrix Chern value \eqref{nrho2} is possible.

For the sake of computation, note that $\rho^{\bm{k}}_{\rm ss}$ is diagonal in the occupation basis $\rho^{\bm{k}}_{\rm ss}=\sum_{n,m\in\{0,1\}} p^{\bm{k}}_{nm}\ket{m,n}\mbox{}_{\bm{k}}\bra{m,n}$, where
\begin{equation*}
\begin{array}{ll}
\ket{00}_{\bm{k}}=\ket{0}\ket{0}, & \ket{10}_{\bm{k}}=\ket{u_{c,\bm{k}}}\ket{0},\\
\ket{01}_{\bm{k}}=\ket{0}\ket{u_{d,\bm{k}}}, & \ket{11}_{\bm{k}}=\ket{u_{c,\bm{k}}}\ket{u_{d,\bm{k}}}.\\
\end{array}
\end{equation*}
The vacuum $\ket{0}$ has no particles and does not depend on  ${\bm{k}}$. If we define the geometric connections for the lower $(c)$ and upper $(d)$ bands:
\begin{equation}
A^{\alpha}_i({\bm{k}}):={\rm i}\bra{u_{\alpha,\bm{k}}} \partial_i u_{\alpha,\bm{k}}\rangle, \quad \alpha = c, d,
\label{bandconnections}
\end{equation}
then it is possible to express the connection $A^\rho_i(\bm{k})$ in terms of the previous ones (see Appendix \ref{app_D}):
\begin{equation}\label{ArhoSimplified5}
A^\rho_i(\bm{k})=\bar{n}_F(E_1^{\bm{k}})A^{c}_i({\bm{k}})+\bar{n}_F(E_2^{\bm{k}})A^{d}_i({\bm{k}}).
\end{equation}
Note that in the $T\rightarrow0$ limit, we recover the standard Berry connection $A^\rho_i(\bm{k}) \rightarrow A^{c}_i({\bm{k}})$, as the steady state \eqref{steady} approaches the Fermi sea with the fully occupied lower band.

\begin{figure}[t]
\centering
\includegraphics[width=0.48\textwidth]{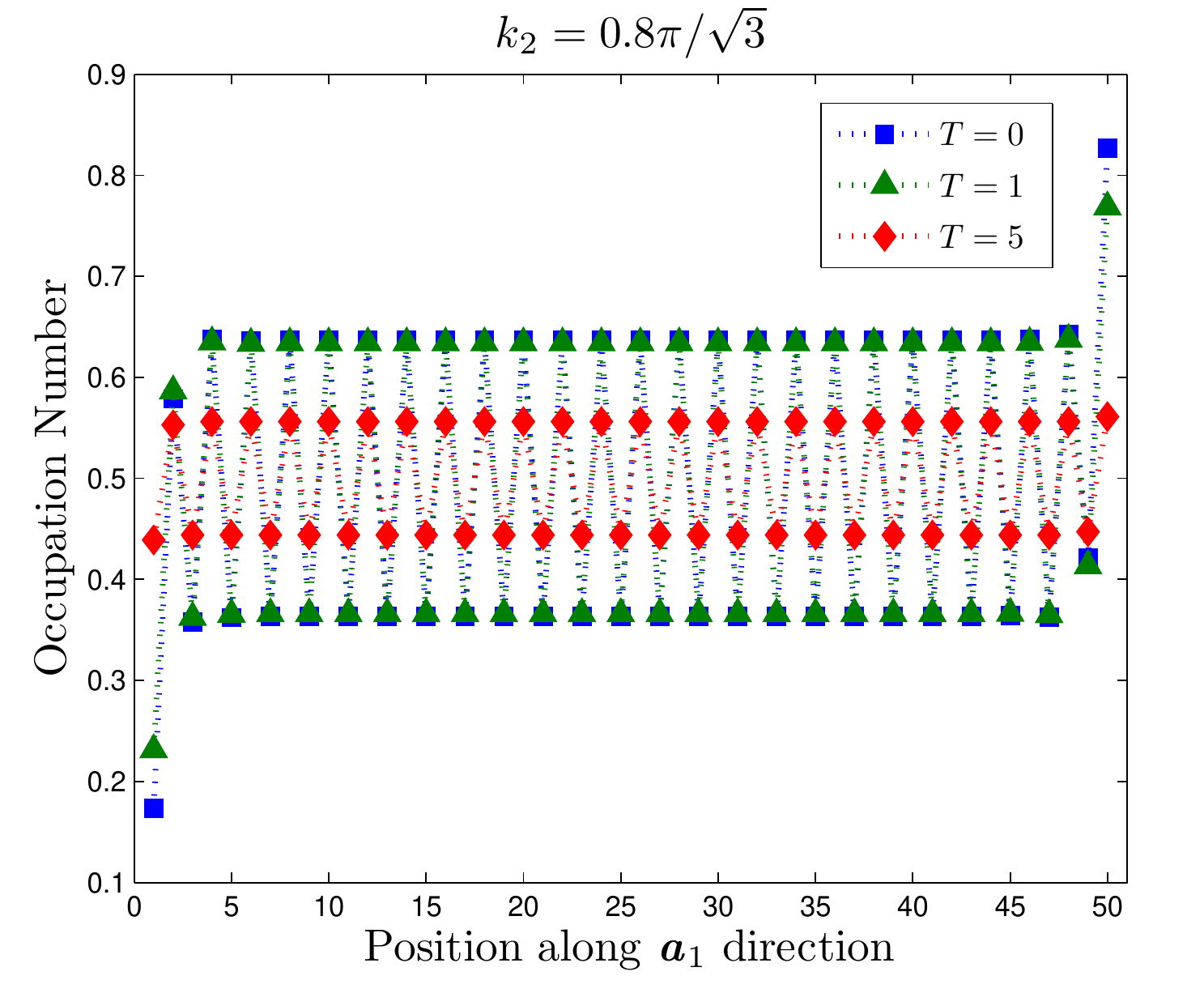}
\caption{Occupation along the direction $\bm{a}_1$ for all particles with momentum $k_2=\frac{4}{5}\frac{\pi}{|\bm{a}_2|}$, $M=0$, and $\phi=\pi/2$), for different values of $T$ (in units of $t_2=1$). Note the presence of edge states at finite temperature [Eq. \eqref{DMedgestates}] in the positions 1 and 50 along the direction $\bm{a}_1$. However, as the temperature $T$ significantly increases, the population of the edge modes becomes similar to the population of the bulk modes.}
\label{plotPopulationT}
\end{figure}

The Chern value can be now computed by integrating the curvature form of $A^\rho_i(\bm{k})$ [or by using the simplified expression \eqref{nrho2}]. The color map in Fig. \ref{PhaseDiagram} represents the Chern value for different values of $M$ and $\phi$ and different bath temperatures. Note its nice properties: it is zero for any choice of $M$ and $\phi$ for all temperatures if it is zero at $T=0$. This manifests that the topological order cannot be created by increasing temperature. Moreover, as $T$ increases, the absolute value of the Chern value decreases, and in the limit of $T\rightarrow\infty$ we obtain ${\rm n}_{\rm Ch}^\rho\rightarrow0$ for all $M$ and $\phi$. This is in agreement with the common intuition that at infinite temperature any kind of order should be spoiled.

\section{Mixed Edge States and Master Equation}
\label{sec:dissipativeedgestates}

A physical signature of a TI phase is the existence of gapless (metallic) edge
states. Thus, once we have mathematically characterized the phase diagram of the Haldane model under dissipation by means of the density matrix Chern value, we wonder about the fate of the chiral edge states of the Haldane model at finite temperature.

To that aim, we consider the Haldane model placed on a cylindrical geometry, where we take periodic boundary conditions just along one spatial dimension, say $\bm{a}_2$. In such a case the momentum $k_2$ along the $\bm{a}_2$ direction is a good quantum number and the Haldane Hamiltonian can be diagonalized obtaining (see Appendix \ref{app_B} for more details)
\begin{equation}
H_{\rm s}=\sum_{k_2\in {\rm B.Z.}}H(k_2)=\sum_{\substack{m\\k_2\in {\rm B.Z.}}} E_{m}^{k_2} f^\dagger_{m,k_2} f_{m,k_2},
\end{equation}
Here, the diagonal modes $f_{m,k_2}$ mix both species of fermions $a_{m,k_2}$ and $b_{m,k_2}$.

By imposing this geometry also in the interaction Hamiltonian \eqref{Hint}, we derive the following dynamical equation for the system (see Appendix \ref{app_C})
\begin{equation}
\begin{split}
\frac{d\rho_{\rm s}(t)}{dt}=&\sum_{k_2\in {\rm B.Z.}}{\cal L}_{k_2}[\rho_{\rm s}(t)]=\sum_{k_2\in {\rm B.Z.} }\bigg(-{\rm i}[H(k_2),\rho_{\rm s}(t)]\\
+&\sum_{m}\Big(\gamma\big(E_m^{k_2}\big)\bar{n}_F\big(E_m^{k_2}\big) \mathcal{D}_{f^{\dagger}_{(m,k_2)}}[\rho_{\rm s}(t)]\\
+&\gamma\big(E_m^{k_2}\big)\big[1-\bar{n}_F\big(E_m^{k_2}\big)\big] \mathcal{D}_{f_{(m,k_2)}}[\rho_{\rm s}(t)]\Big)\bigg),
\end{split}
\label{masterEqCylinder}
\end{equation}
where
\begin{equation}
\mathcal{D}_{K}[\rho_{\rm s}(t)]:=K\rho_{\rm s}(t)K^\dagger-\frac{1}{2}\{K^{\dagger}K,\rho_{\rm s}(t)\}.
\end{equation}

Again, the Gibbs state at the same temperature as the bath is the unique steady state of Eq. \eqref{masterEqCylinder},
\begin{equation}\label{steadyCylinder}
\rho_{\beta}=\frac{\text{e}^{-\beta \sum_{k_2}H(k_2)}}{Z}=\bigotimes_{k_2} \frac{\text{e}^{-\beta H(k_2)}}{Z_{k_2}},
\end{equation}
with  $Z_{k_2}=\traza \left[\text{e}^{-\beta H(k_2)}\right].$
Therefore, as long as the values of $M$, $t_2$ and $\phi$ are such that the system exhibits symmetry protected topological order (see Fig. \ref{PhaseDiagram}), two of the modes which diagonalize each $H(k_2)$, say $f_{(L,k_2)}$ and $f_{(R,k_2)}$, correspond to edge states and the Gibbs state is a tensor product in $k_2$ of states of the form
\begin{equation}\label{steadyCylinder2}
\rho_\beta(k_2)=\frac{{\rm e}^{-\beta H(k_2)}}{Z_{k_2}}=\rho_\beta^{\rm L}(k_2)\otimes \rho_\beta^{\rm bulk}(k_2)\otimes \rho_\beta^{\rm R}(k_2),
\end{equation}
where
\begin{equation}\label{DMedgestates}
\rho_\beta^{\rm L,R}(k_2):=\frac{{\rm e}^{-\beta E_{L,R}(k_2)f^\dagger_{(L,R,k_2)}f_{(L,R,k_2)}}}{1+{\rm e}^{-\beta E_{L,R}(k_2)}},
\end{equation}
are Gibbs states of the edge modes. However, as temperature increases (see again Fig. \ref{PhaseDiagram}), the edge modes in the steady state of the Liouvillian \eqref{masterEqCylinder} become delocalized along the transverse direction to the edges. Then, the components $\rho^{\rm L,R}(k_2)$ approach to a maximally mixed state with vacuum $|0\rangle$. In such a situation, there is not a way to distinguish the edge from the bulk modes, because for all $k_2$ the occupation along the direction $\bm{a}_1$ becomes the same and equal to $1/2$. We illustrate this behavior in Fig. \ref{plotPopulationT}.

It is worth stressing that in order to study the dissipative effects on the cylindrical Haldane system we must determine how the boundary conditions of the system affect the dissipator operator. To that aim we need to specify how the dissipator is generated (constructed). In our case this is the result of the weak interaction of the system with local baths. However there are also possible scenarios where the dissipator is the effective result of other external interactions (see for example \cite{Wolf,Baranov,Sebastian,Markus,Geza,Eisert}). Since the process of ``opening'' or ``closing'' a system has a physical meaning, we stress that we need to know how the dissipator is physically generated to obtain its ``open'' and/or ``closed'' counterpart. Note that a dissipator with periodic boundary conditions, may split in different and nonequivalent dissipators once the system is opened along some direction if it is generated in different ways.

\begin{figure}[t]
\includegraphics[width= 0.48\textwidth]{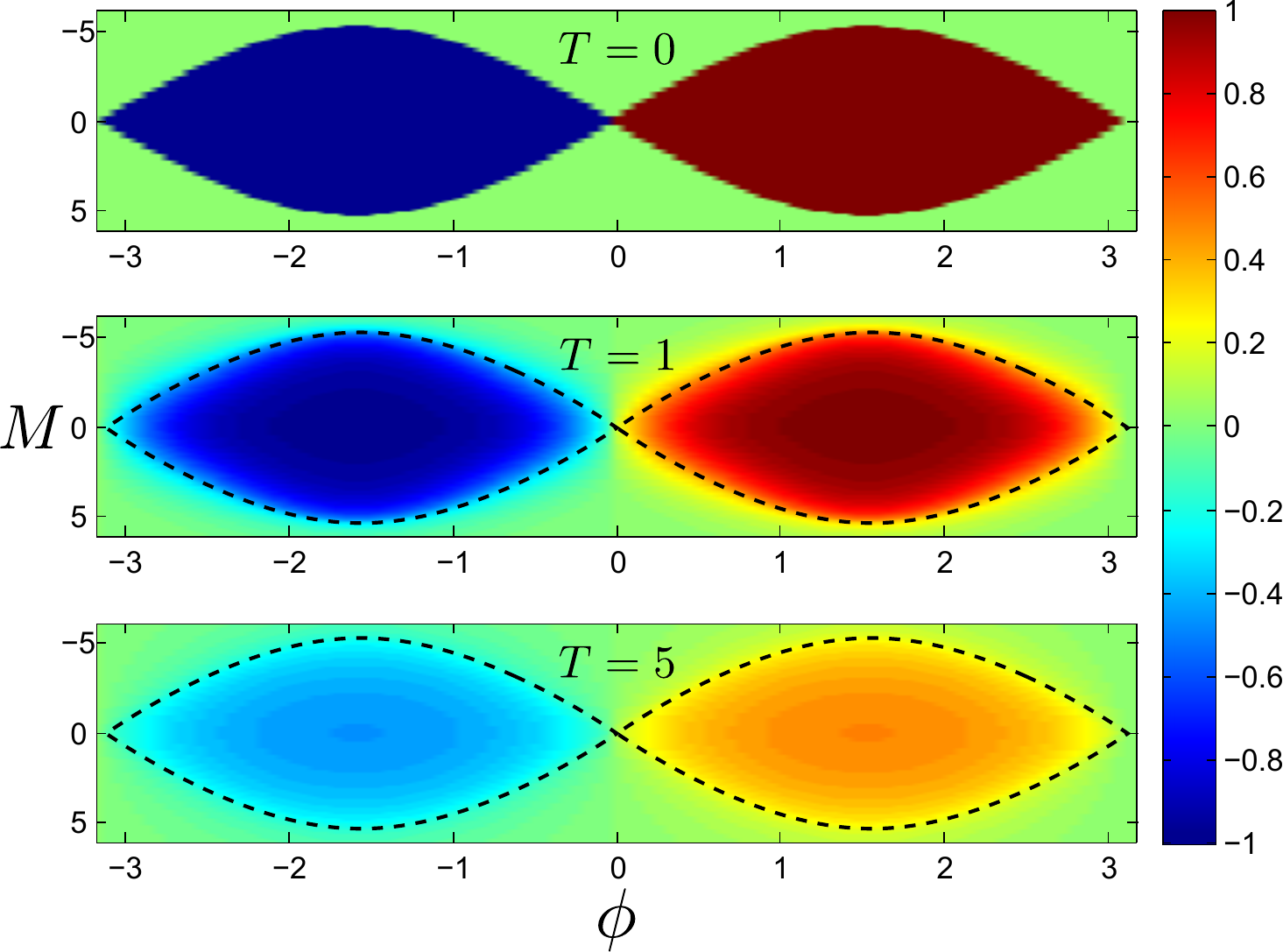}
\caption{Colormap depicting the conductivity Eq. \eqref{conductivity1} for different values of $\phi$, $M$ and the bath temperature $T$ (in units of $t_2=1$). As $T$ increases the conductivity decreases (in absolute value), and for $T=0$ we recover the Chern number result, Fig. \ref{PhaseDiagram}. The dashed black lines enclose the region displaying topological order at $T=0$. Thus, contrarily to the Chern value, the conductivity is not a topological property for finite $T$, as it does not vanish for every point outside this region for any $T$.}
\label{ConductivityPlot}
\end{figure}

\section{Quantum Hall Conductivity and Chern Value at Finite Temperature} \label{sec:conductivity}

We can obtain further physical meaning and implications for the density matrix Chern value \eqref{nrho2} by studying the (quantum Hall) transverse conductivity $\sigma_{xy}$ and its relation to the thermal edge states obtained for the Haldane model. Using the Kubo formula \cite{Kubo_57}  in  linear response theory, it is possible to derive an expression for the transverse Hall conductivity at finite temperature \cite{Xiao_09}:
\begin{equation}
\sigma^{\rho}_{xy}=\frac{e^2}{2\pi h}\sum_{\alpha}\int_{{\rm T}^2}\bar{n}_F(E_{\alpha}^{\bm{k}})F_{xy}^{\alpha}({\bm k})d^2{\bm k}.
\label{conductivity1}
\end{equation}

Note that this expression \cite{comment} is different from the one obtained for the Chern value \eqref{nrho2}. Indeed, the conductivity is not topological at finite temperature, as shown in Fig.~\ref{ConductivityPlot} where non-zero Hall conductivity appears in regions outside the topological regime, in contrast with Fig.~\ref{PhaseDiagram}.  Nonetheless, both quantities can be related by means of the following equation:
\begin{align}\label{conductivity2}
\sigma^{\rho}_{xy}=\frac{e^2}{h}{\rm n}_{\rm Ch}^\rho &+\frac{e^2}{2\pi h}\sum_{\alpha}\int_{{\rm T}^2}\Big\{[\partial_y\bar{n}_F(E_{\alpha}^{\bm{k}})]A^{\alpha}_x(\bm{k})\nonumber\\
&-[\partial_x\bar{n}_F(E_{\alpha}^{\bm{k}})]A^{\alpha}_y(\bm{k})\Big\}d^2{\bm k}.
\end{align}

The second term on the right hand side of \eqref{conductivity2} is the same one that appears for the transverse conductivity of a normal insulator with an applied magnetic field (or a pseudo-magnetic field as for the Haldane model) at $T\not=0$. It corresponds to the conduction by thermal activation of excited electrons in the bulk. For instance, for parameters $t_1=4, t_2=1, \phi=\frac{\pi}{2}$ and $M=6$ in the Haldane model, this term is the only non-zero contribution to the conductivity, as the system is outside the topological regime, and so $ {\rm n}^\rho_{\rm Ch}=0$.

Notwithstanding, the first term on the right hand side of \eqref{conductivity2}, which is nothing but the Chern value previously defined, represents a contribution due to the topological nature of our system and the presence of conducting edge states. For parameters $t_1=4, t_2=1, \phi=\frac{\pi}{2}$ and $M=0$ in the Haldane model, the system is within the topological regime and this new term shows up. Note that at $T=0$ we recover the well known TKNN expression for the conductivity:
\begin{equation}
\sigma^{\rho}_{xy}\xrightarrow[T\longrightarrow0]{}\frac{e^2}{h}~{\nu}_{\rm Ch},
\label{cond_lim}
\end{equation}\\
where $\nu_{\rm Ch}$ denotes the standard Chern number.

\section{Conclusions}
\label{sec:conclusions}

We have studied topological insulating phases in the presence of dissipation. After introducing the notion of band Liouvillian, we address the characterization of the topological order of its steady states by resorting to the density matrix Chern value, a topological indicator that is an extension of the Chern number for pure states. The Haldane model of a 2D TI in contact with a thermal bath offers a nice testbed to study these phenomena. More concretely, we compute phase diagrams at finite temperature based on the Chern value, and corroborate that topological order decreases as the bath temperature increases. Thus, from a topologically disordered state it is not possible to induce a topologically ordered phase just by warming the system. However, a topologically ordered state may remain ordered at finite temperature $T$ except at the limit $T\rightarrow \infty$. This has to be compared with the previous study \cite{Creutz_Ladder2012} where symmetry-protected topological order turned out to be lost for a dissipative system in the presence of noise not generated by a band Liouvillian. Our results may also have direct application in recent studies regarding dissipation on Majorana fermions in topological superconductors \cite{Baranov,Sebastian,Mazza}.

Complementarily, we study the properties of the Haldane model coupled to a thermal bath under cylindrical boundary conditions. We find that the the steady state splits in three different, generally mixed, substates [Eq. \eqref{steadyCylinder2}]. Two of them are associated with creation or annihilation of fermionic gapless edge modes, and the other one accounts for the same process just in the bulk modes. In the limit $T\rightarrow0$, we recover the properties of the usual Haldane model.

Finally, we examine the relation between the density-matrix Chern value and the conductivity at finite temperature. We show that the latter is a topological property, in contrast to the Chern value. This fact is due to the presence of an extra term which accounts for the conductivity generated by thermal activation of electrons in the bulk, which has not a particular topological meaning and is present in normal insulators. In this regard, provided that the gap between conduction and valence bands is large enough, the Chern value may be approximately estimated by measuring the conductivity, as the thermal activation would be very small. However, recent results \cite{Bloch}, suggest that a direct measurement of the density matrix Chern value could be possible in optical lattice realizations.

\acknowledgements
We thank Alexandre Dauphin for fruitful discussions. This
work has been supported by the Spanish MINECO grant FIS2012-33152,
CAM research consortium QUITEMAD S2009-ESP-1594, European Commission
PICC: FP7 2007-2013, Grant No.~249958, UCM-BS grant GICC-910758.

\appendix

\section{Gauge Invariance of the Density Matrix Chern Value}
\label{app_A}

In this appendix we provide a proof that the density matrix Chern value, Eq. \eqref{nrho}, is fully gauge invariant with respect to ${\rm U}(1)^N$ transformations of the mixed state. To that aim, consider a ${\rm U}(1)^N$ gauge transformation on the eigenstates of $\rho_{\rm ss}^{\bm{k}}=\sum_{\alpha=1}^{N}p_\alpha^{\bm{k}}\ket{\psi_{\alpha,\bm{k}}}\bra{\psi_{\alpha,\bm{k}}}$:
\begin{equation}
\ket{\psi_{\alpha,\bm{k}}}~\longrightarrow~\ket{\tilde{\psi}_{\alpha,\bm{k}}}={\rm e}^{{\rm i}\phi_{\bm k}^{\alpha}}\ket{\psi_{\alpha,\bm{k}}}.
\label{gtr}
\end{equation}
The Berry-type connection Eq. \eqref{ArhoSimplified2} and the associated purified curvature \eqref{Fdef} transform as:
\begin{align}
\tilde{A}^{\rho}_i({\bm{k}})&=A^{\rho}_i({\bm{k}}) -\sum_{\alpha}p_\alpha^{\bm{k}}\partial_i\phi_{\bm k}^{\alpha}, \\
\tilde{F}_{ij}^\rho(\bm{k})&=F_{ij}^\rho(\bm{k})+\sum_{\alpha}\partial_j(p_\alpha^{\bm{k}}\partial_i\phi_{\bm k}^{\alpha})-\partial_i(p_\alpha^{\bm{k}}\partial_j\phi_{\bm k}^{\alpha})
\label{Arho2}
\end{align}

On the other hand, the translational invariance of the lattice imposes that $\rho_{\rm ss}^{\bm{k}}=\rho_{\rm ss}^{\bm{k}+\bm{G}}$, where ${\bm G}$ is a vector in the reciprocal lattice. Then, since the eigenbasis of $\rho_{\rm ss}^{\bm{k}}$ is single-valued, we conclude that
\begin{equation}
\ket{\psi_{\alpha,\bm{k}}}=\ket{\psi_{\alpha,\bm{k + G}}}
\end{equation}
independently of the gauge. This implies
\begin{equation}
\ket{\tilde{\psi}_{\alpha,\bm{k}}}=\ket{\tilde{\psi}_{\alpha,\bm{k + G}}}\Rightarrow{\rm e}^{{\rm i}\phi_{\bm k}^{\alpha}}\ket{\psi_{\alpha,\bm{k}}}={\rm e}^{{\rm i}\phi_{\bm{ k+G}}^{\alpha}}\ket{\psi_{\alpha,\bm{k+G}}}
\label{invtr2}
\end{equation}
Thus, we obtain that the gauge phase satisfies the relation
\begin{equation}
\phi_{\bm k}^{\alpha}=\phi_{\bm{k+G}}^{\alpha}~~(\text{mod}~2\pi).
\label{phi}
\end{equation}

The Chern value (in 2D for simplicity)  is given by
\begin{equation}
{\rm n}_{\rm Ch}^\rho=\frac{1}{2\pi}\int_{{\rm T}^2}F_{12}^{\rho}d^2k,
\label{ch_val}
\end{equation}

Performing a ${\rm U}(1)^N$ gauge transformation and using \eqref{Arho2},
\begin{widetext}
\begin{eqnarray}
\tilde{{\rm n}}_{\rm Ch}^\rho&=&\frac{1}{2\pi}\int_{{\rm T}^2}\tilde{F}_{12}^{\rho}d^2k={\rm n}_{\rm Ch}^\rho+\frac{1}{2\pi}\sum_{\alpha}\int^{\pi}_{-\pi}dk_1\Big[\int^{\pi}_{-\pi}dk_2\partial_{k_2}(p_\alpha^{\bm{k}}\partial_{k_1}\phi_{\bm k}^{\alpha})\Big]-\frac{1}{2\pi}\int^{\pi}_{-\pi}{dk_2}\Big[\int^{\pi}_{-\pi}{dk_1}\partial_{k_1}(p_\alpha^{\bm{k}}\partial_{k_2}\phi_{\bm k}^{\alpha})\Big]\nonumber\\
&=&{\rm n}_{\rm Ch}^\rho+\frac{1}{2\pi}\sum_{\alpha}\int^{\pi}_{-\pi}{dk_1}\Big[p_\alpha(k_1,k_2=\pi)\partial_{k_1}\phi^{\alpha}(k_1,k_2=\pi)-p_\alpha(k_1,k_2=-\pi)\partial_{k_1}\phi^{\alpha}(k_1,k_2=-\pi)\Big]\nonumber\\
&-&\frac{1}{2\pi}\sum_{\alpha}\int^{\pi}_{-\pi}{dk_2}\Big[p_\alpha(k_1=\pi,k_2)\partial_{k_2}\phi^{\alpha}(k_1=\pi,k_2)-p_\alpha(k_1=-\pi,k_2)\partial_{k_2}\phi^{\alpha}(k_1=-\pi,k_2)\Big],
\label{ch_val}
\end{eqnarray}
where $k_1$ and $k_2$ are the two periodic directions along the 2-torus. The weights $p_\alpha^{\bm{k}}$ are periodic in the B.Z. In particular for the Gibbs' state, these are functions of the energies of the system. Thus, we have $p_\alpha^{\bm{k}}=p_\alpha^{\bm{k+G}}$ and then, it follows that
\begin{eqnarray}
\tilde{{\rm n}}_{\rm Ch}^\rho={\rm n}_{\rm Ch}^\rho&+&\frac{1}{2\pi}\sum_{\alpha}\int^{\pi}_{-\pi}{dk_1}\Big\{p_\alpha(k_1,k_2=\pi)\partial_{k_1}\big[\phi^{\alpha}(k_1,k_2=\pi)-\phi^{\alpha}(k_1,k_2=-\pi)\big]\Big\}\nonumber\\
&-&\frac{1}{2\pi}\sum_{\alpha}\int^{\pi}_{-\pi}{dk_2}\Big\{p_\alpha(k_1=\pi,k_2)\partial_{k_2}\big[\phi^{\alpha}(k_1=\pi,k_2)-\phi^{\alpha}(k_1=-\pi,k_2)\big]\Big\}.
\label{ch_val2}
\end{eqnarray}
\end{widetext}
At this point, we make use of Eq. \eqref{phi} and thus
\begin{equation}
\partial_{k_{x,y}}\bigg(\phi_{\bm k}^{\alpha}-\phi_{\bm{k+G}}^{\alpha}\bigg)=0.
\label{phi3}
\end{equation}
Hence, we can further simplify \eqref{ch_val2} using \eqref{phi3}, arriving at the fundamental result
\begin{equation}
\tilde{{\rm n}}_{\rm Ch}^\rho={\rm n}_{\rm Ch}^\rho.
\label{ch_val3}
\end{equation}

To summarize, the purified Berry curvature is only U(1) gauge invariant, however, the Chern value is ${\rm U}(1)^N$ gauge invariant and consequently it can represent a physical observable.

\section{Haldane Model Geometries}
\label{app_B}
\begin{figure}[t]
\centering
\includegraphics[width=0.4\textwidth]{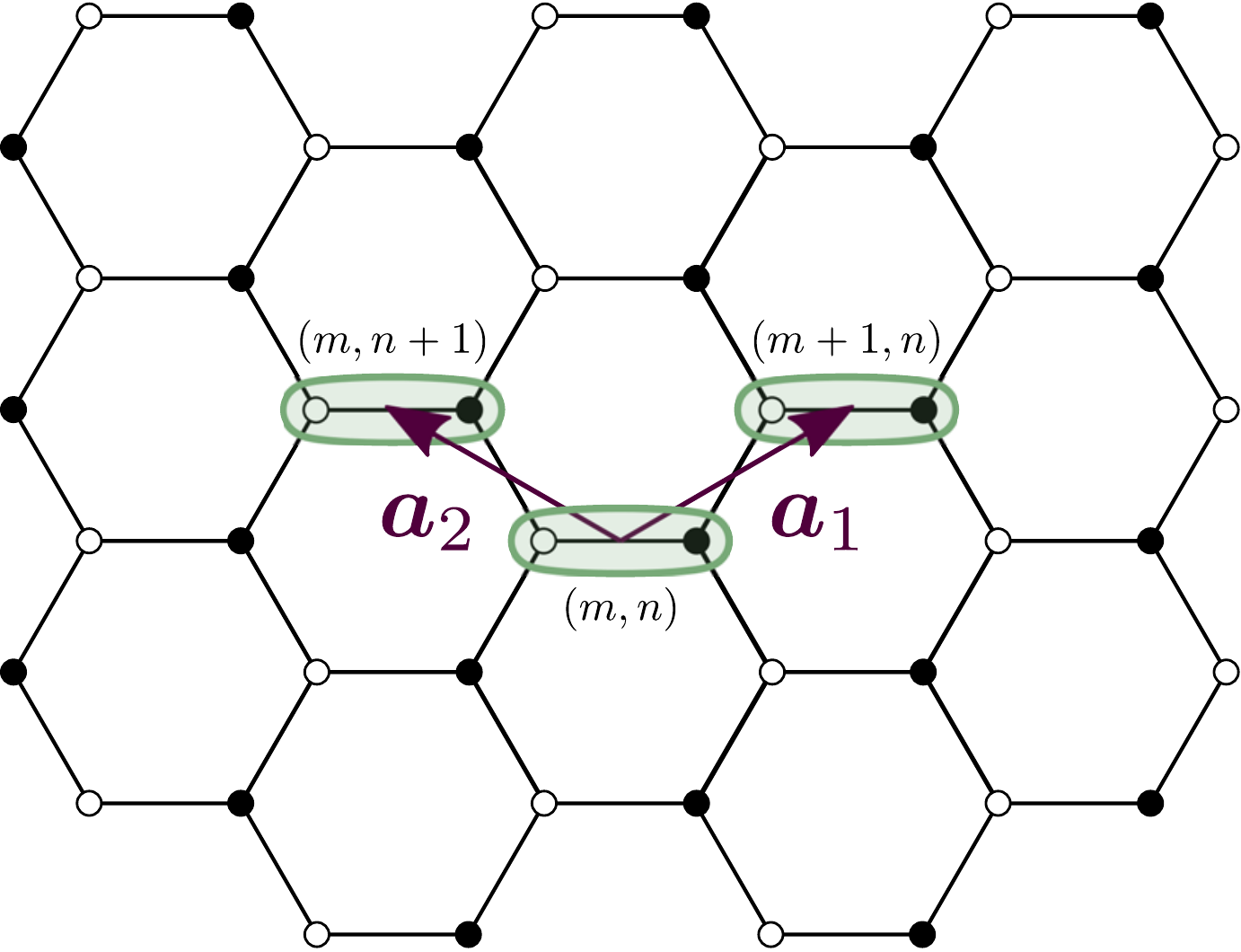}
\caption{System of coordinates $\{\bm{a}_1,\bm{a}_2\}$ taken to write the Haldane Hamiltonian in real space \eqref{HsOpen} . The solid white and black circles denote fermions $a$ and $b$, respectively, and the green enclosures highlight the two-site unit cell.}
\label{plotAMMA}
\end{figure}

In order to write explicitly the Haldane Hamiltonian \cite{Haldane} in real space we will consider the system of coordinates $\{\bm{a}_1,\bm{a}_2\}$ represented in Fig. \ref{plotAMMA}, with $\bm{a}_1=\frac{1}{2}(3,\sqrt{3})$ and $\bm{a}_2=\frac{1}{2}(-3,\sqrt{3})$ for lattice spacing $a=1$. We write $a_{(m,n)}$ ($b_{(m,n)}$) for the fermionic operator of kind $a$ ($b$) in the position $\bm{r}_{(m,n)}=m\bm{a}_1+n\bm{a}_2$. Then, the Haldane Hamiltonian in real space reads as
\begin{align}\label{HsOpen}
H_{\rm s}&:=\sum_{m,n}\bigg( \tfrac{M}{2}[a^\dagger_{(m,n)}a_{(m,n)}-b^\dagger_{(m,n)}b_{(m,n)}]\nonumber \\
&+ t_1\Big[a^\dagger_{(m,n)}b_{(m,n)}+a^\dagger_{(m+1,n)}b_{(m,n)}+a^\dagger_{(m,n)}b_{(m,n+1)}\Big] \nonumber \\
&+ t_2\Big[{\rm e}^{{\rm i}\phi} a^\dagger_{(m,n)}a_{(m+1,n+1)} + {\rm e}^{-{\rm i}\phi} a^\dagger_{(m,n)}a_{(m+1,n)} \nonumber\\
&+ {\rm e}^{-{\rm i}\phi} a^\dagger_{(m,n)}a_{(m,n+1)}+{\rm e}^{-{\rm i}\phi} b^\dagger_{(m,n)}b_{(m+1,n+1)} \nonumber \\
&+ {\rm e}^{{\rm i}\phi} b^\dagger_{(m,n)}b_{(m+1,n)}+{\rm e}^{{\rm i}\phi} b^\dagger_{(m,n)}b_{(m,n+1)}\Big]+{\rm h.c.}\bigg).
\end{align}

\subsection{Toroidal geometry}
By taking periodic boundary conditions in both spatial directions, we may write the Hamiltonian \eqref{HsOpen} in the reciprocal space using the Fourier-transformed operators
\begin{eqnarray}
a_{(n,m)}=\frac{1}{\sqrt{N}}\sum_{\bm{k}\in {\rm B.Z.}}{\rm e}^{{\rm i}\bm{k}\cdot\bm{r}_{(m,n)}}a_{\bm k},\\
b_{(n,m)}=\frac{1}{\sqrt{N}}\sum_{\bm{k}\in {\rm B.Z.}}{\rm e}^{{\rm i}\bm{k}\cdot\bm{r}_{(m,n)}}b_{\bm k},
\end{eqnarray}
where B.Z. stands for Brillouin zone which is a hexagon with vertices in the ${\bm k}=(k_1,k_2)$ points
\begin{eqnarray}
&\left(0,\frac{4\pi}{3\sqrt{3}}\right), \left(\frac{2\pi}{3},\frac{2\pi}{3\sqrt{3}}\right), \left(\frac{2\pi}{3},-\frac{2\pi}{3\sqrt{3}}\right),\\
&\left(0,-\frac{4\pi}{3\sqrt{3}}\right), \left(-\frac{2\pi}{3},-\frac{2\pi}{3\sqrt{3}}\right), \left(-\frac{2\pi}{3},\frac{2\pi}{3\sqrt{3}}\right),
\end{eqnarray}
and $N$ is the total number of two-site unit cells. Thus the Haldane Hamiltonian is rewritten as
\begin{equation}
H_{\rm s}=\sum_{\bm k}(a^{\dagger}_{\bm k}~,~b^{\dagger}_{\bm k})H({\bm k})\begin{pmatrix} a_{\bm k}\\ b_{\bm k} \end{pmatrix}.
\label{HsA}
\end{equation}
Here,
\begin{eqnarray}
H_{11}({\bm k}) &=& M+2t_2\sum_i\cos[\phi+({\bm k}\cdot {\bm b}_i)], \nonumber\\
H_{12}({\bm k}) &=&  H({\bm k})_{21}^{\ast} = t_1\sum_i\text{e}^{-{\rm i}{\bm k}\cdot {\bm a}_i}, \\
H_{22}({\bm k}) &=&  -M+2t_2\sum_i\cos[\phi-({\bm k}\cdot {\bm b}_i)] \nonumber
\end{eqnarray}
with
\begin{eqnarray}
{\bm b}_1 &=& -\left(3,\sqrt{3}\right)/2, \nonumber\\
{\bm b}_2 &=&\left(3,-\sqrt{3}\right)/2, \\
{\bm b}_3 &=& \left(0,\sqrt{3}\right).\nonumber
\end{eqnarray}

By diagonalizing the matrix $H(\bm{k})$ we obtain
\begin{equation}
H_{\rm s}=\sum_{\bm{k}}E_1^{\bm{k}}c^{\dagger}_{\bm{k}}c_{\bm{k}} + E_2^{\bm{k}}d^{\dagger}_{\bm{k}}d_{\bm{k}},
\label{Hsdiag}
\end{equation}
where the eigenvalues are given by
\begin{equation}
E^{\bm k}_{1,2}=2t_2(\cos{\phi})~\xi_1(\bm{k})\mp\sqrt{ \Delta(\bm{k})}
\end{equation}
with
\begin{multline}
\Delta(\bm{k}) := M^2 + t_1^2[3+2\xi_1(\bm{k})]\\
+4t_2^2(\sin^2 \phi)~[\xi_2(\bm{k})]^2 -4Mt_2(\sin{\phi})\xi_2(\bm{k}),
\end{multline}
and
\begin{equation}
\begin{array}{l}
\xi_1(\bm{k}) :=\sum_i\cos{{\bm k}\cdot {\bm b}_i}=2\cos{\tfrac{3k_1}{2}}\cos{\tfrac{\sqrt{3}k_2}{2}}+\cos{\sqrt{3}k_2}, \\
\xi_2(\bm{k}) :=\sum_i\sin{{\bm k}\cdot {\bm b}_i}=-2\cos{\tfrac{3k_1}{2}}\sin{\tfrac{\sqrt{3}k_2}{2}}+\sin{\sqrt{3}k_2}.
\end{array}
\end{equation}
The operators $c_{\bm{k}}$ and $d_{\bm{k}}$ are related to $a_{\bm{k}}$ and $b_{\bm{k}}$ via the canonical transformation
\begin{equation}
\begin{pmatrix} a_{\bm{k}}\\ b_{\bm{k}} \end{pmatrix}=\frac{1}{\sqrt{1+|x_{\bm{k}}|^2}}\begin{pmatrix} x_{\bm{k}} & 1\\ -1 & x^{\ast}_{\bm{k}} \end{pmatrix}\begin{pmatrix} c_{\bm{k}}\\ d_{\bm{k}} \end{pmatrix},
\end{equation}
where
\begin{equation}
x_{\bm k}:=\frac{t_1\sum_i\text{e}^{-{\rm i}{\bm k}\cdot {\bm a}_i}}{M-2t_2(\sin{\phi})~\xi_2(\bm{k})+\sqrt{\Delta(\bm{k})}}.
\end{equation}

\subsection{Cylindrical geometry}

\begin{figure}[t]
\centering
\includegraphics[width=0.45\textwidth]{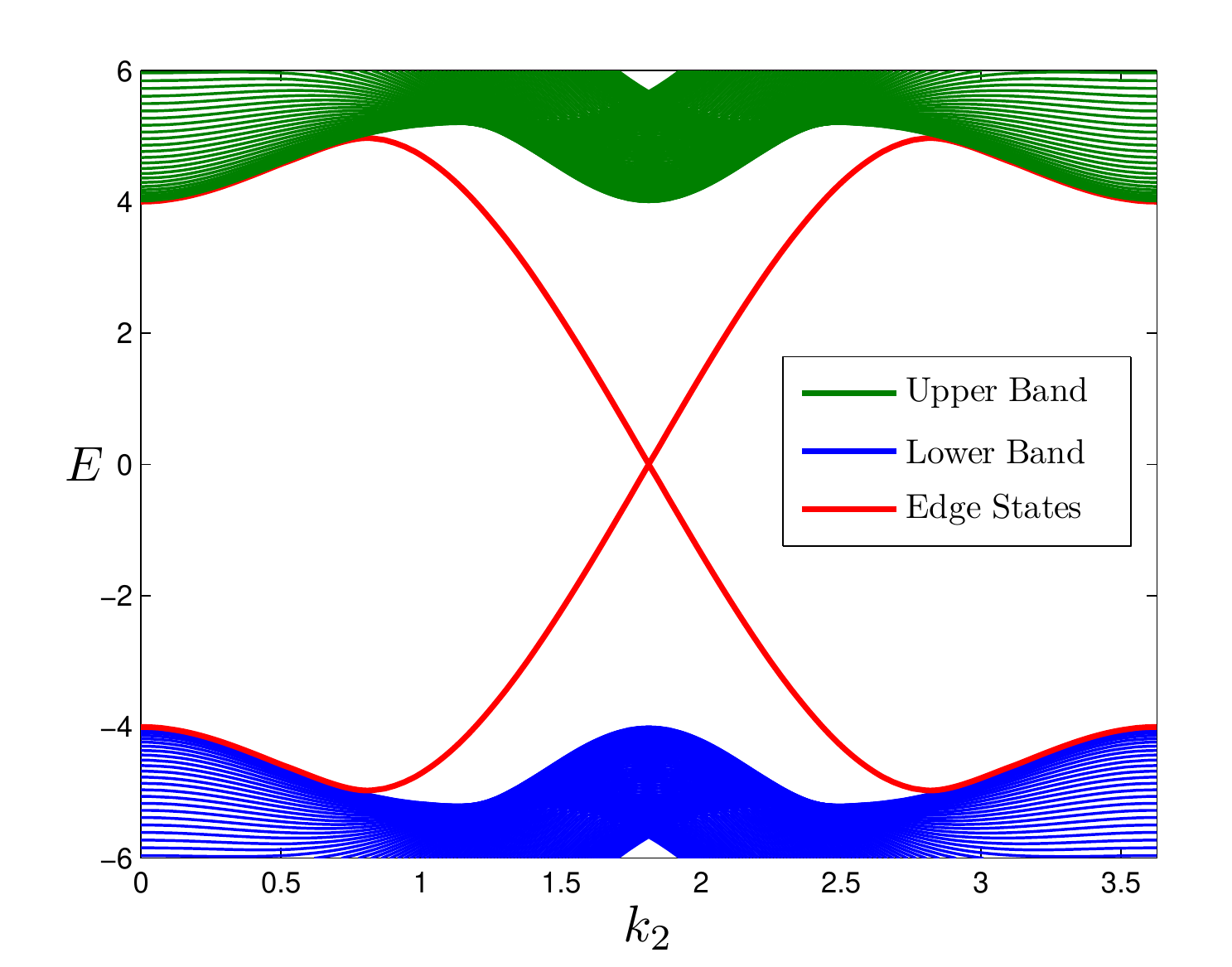}
\caption{Energy bands of the Hamiltonian \eqref{HsCilinder} as a function of the momentum $k_2$. The red lines correspond to the edge state modes. The rest of parameters are $M=0$, $\phi=\pi/2$ and $t_1=4$ (in units of $t_2=1$).}
\label{plotEdgeBands}
\end{figure}

In this case we take periodic boundary conditions along the direction $\bm{a}_2$ and open boundaries along $\bm{a}_1$, so that we work with a cylindrical configuration. The inverse Fourier transform along the $\bm{a}_2$ direction of the fermionic operators is given by
\begin{eqnarray}
a_{(m,n)}&=&\frac{1}{\sqrt{N}_2}\sum_{k_2\in{\rm B.Z.}}{\rm e}^{{\rm i} k_2 n |\bm{a}_2|}a_{(m,k_2)},\\
b_{(m,n)}&=&\frac{1}{\sqrt{N}_2}\sum_{k_2\in{\rm B.Z.}}{\rm e}^{{\rm i} k_2 n |\bm{a}_2|}b_{(m,k_2)},
\end{eqnarray}
where the Brillouin zone corresponds to the interval $k_2\in(-\pi/|\bm{a}_2|,\pi/|\bm{a}_2|)=(-\pi/\sqrt{3},\pi/\sqrt{3})$, and $N_2$ is the number of two-site basic cells along the direction $\bm{a}_2$. By using these equations in the Hamiltonian \eqref{HsOpen} we obtain
\begin{align}\label{HsCilinder}
H_{\rm s}&=\sum_{k_2\in{\rm B.Z}}\sum_{m} \bigg(\Big[\tfrac{M}{2}+t_2\cos\big(\sqrt{3}k_2-\phi\big)\Big]a^\dagger_{(m,k_2)}a_{(m,k_2)} \nonumber \\
&+\left[-\tfrac{M}{2}+t_2\cos(\sqrt{3}k_2+\phi)\right]b^\dagger_{(m,k_2)}b_{(m,k_2)} \nonumber \\
&+t_1\left[ a^\dagger_{(m+1,k_2)}b_{(m,k_2)}+\left(1+{\rm e}^{{\rm i}\sqrt{3}k_2}\right)a^\dagger_{(m,k_2)}b_{(m,k_2)}  \right] \nonumber\\
&+t_2\Big[ \left({\rm e}^{{\rm i}(\sqrt{3}k_2+\phi)}+{\rm e}^{-{\rm i}\phi}\right) a^\dagger_{(m,k_2)}a_{(m+1,k_2)} \nonumber \\
&+ \left({\rm e}^{{\rm i}(\sqrt{3}k_2-\phi)}+{\rm e}^{{\rm i}\phi}\right) b^\dagger_{(m,k_2)}b_{(m+1,k_2)} \Big]+{\rm h.c.}\bigg).
\end{align}
This Hamiltonian has the structure $H_{\rm s}=\sum_{k_2\in {\rm B.Z.} } H(k_2)$, therefore we can diagonalize $H$ by diagonalizing each $H(k_2)$. In Fig. \ref{plotEdgeBands} we have depicted the behavior of the eigenvalues of $H(k_2)$ as a function of $k_2$. The red lines connecting the upper and lower bands correspond to the edge state modes, which are localized on the edges of the direction $\bm{a}_1$.

\section{Derivation of the Master Equation for the Haldane Model}
\label{app_C}

In this section we derive the dynamical equation (master equation) for the Haldane model coupled to a thermal bath. We assume the usual condition of weak system-bath coupling, which is a standard assumption for thermalization.

The total Hamiltonian of the problem considered reads as follows

\begin{equation}
H:=H_{\rm s}+H_{\rm b}+H_{\rm s-b}.
\label{H}
\end{equation}
The first term, $H_{\rm s}$, is the Haldane Hamiltonian \eqref{HsOpen}. The second term in \eqref{H}, $H_{\rm b}$, is the free Hamiltonian of the local baths,
\begin{equation}
H_{\rm b}:=\sum_{i,\bm{r}}\epsilon^i\left(A^{i\dagger}_{\bm{r}}A^i_{\bm{r}}+B^{i\dagger}_{\bm{r}}B^i_{\bm{r}}\right),
\end{equation}
where $A$ and $B$ stand for independent fermionic bath operators that satisfy the canonical anti-commutation relations
$\{A^i_n,A^{j\dagger}_{n'}\}=\delta _{n,n'}\delta_{i,j}~,~\{A^i_n,A^{j}_{n'}\}=0$ and analogously for $B^i_n$. The index $\bm{r}$ denotes the position of the local bath on the lattice and $i$ runs over the bath degrees of freedom. Also, $\epsilon^i$ represents the energy of each mode $i$ of the bath which is assumed to be independent of the lattice site.
Finally, the third term in \eqref{H}, $H_{\rm s-b}$, is given by
\begin{equation} \label{HintSM}
H_{\rm s-b}:=\sum_{i,\bm{r}}g^i(a^{\dagger}_{\bm{r}}\otimes A^{i}_{\bm{r}}+a_{\bm{r}}\otimes A^{i{\dagger}}_{\bm{r}}+b^{\dagger}_{\bm{r}}\otimes B^{i}_{\bm{r}}+b_{\bm{r}}\otimes B^{i{\dagger}}_{\bm{r}}),
\end{equation}
and describes an exchange of fermions between system and bath mediated by a coupling constant $g^i$ which may depend on the specific mode $i$ of the baths.

The total dynamics of system and bath is given by the Liouville-Von-Neumann equation:
\begin{equation}
\frac{d\rho}{dt} =-{\rm i}[H,\rho].
\label{ME}
\end{equation}
After taking the interaction picture with respect to $H_0=H_{\rm s}+H_{\rm b}$,
\begin{equation}
\frac{d\tilde{\rho}}{dt} =-{\rm i}[\tilde{H}_{\rm s-b},\tilde{\rho}] \quad \text{with}\ \begin{cases}
\tilde{\rho}:=\text{e}^{{\rm i}H_0t}\rho\text{e}^{-{\rm i}H_0t},\\
\tilde{H}_{\rm s-b}:=\text{e}^{{\rm i}H_0t}H_{\rm s-b}\text{e}^{-{\rm i}H_0t}.
\end{cases}
\label{ipicture}
\end{equation}

For small $\tilde{H}_{\rm s-b}$, the system dynamics is approximately given (see \cite{Alicki,GZ,BrPe,Libro}) by the equation
\begin{equation}\label{weakCoupling}
\frac{d\tilde{\rho}_{\rm s}}{dt} = -\int^{\infty}_0ds~{\rm Tr_{\rm b}}[\tilde{H}_{\rm s-b}(t),[\tilde{H}_{\rm s-b}(t-s),\tilde{\rho}_{\rm s}(t)\otimes\rho^\beta_{\rm b}]],
\end{equation}
where ${\rm Tr_{\rm b}}$ denotes the trace over the bath degrees of freedom and $\rho_{\rm b}^\beta$ is the initial state of the bath, which we assumed to be the Gibbs state
\begin{equation}
\rho^\beta_{\rm b}:=\frac{{\rm e} ^{-\beta H_{\rm b}} }{Z}.
\end{equation}

\subsection{Toroidal geometry}
We consider periodic boundary conditions and take Fourier transforms in the Hamiltonian \eqref{H}. For the first term we obtain \eqref{HsA}, for the second we have
\begin{equation}
H_{\rm b}=\sum_{i,\bm{k}}\epsilon^i\left(A^{i\dagger}_{\bm{k}}A^i_{\bm{k}}+B^{i\dagger}_{\bm{k}}B^i_{\bm{k}}\right),
\end{equation}
and finally for the interaction term
\begin{equation}\label{Hintk}
H_{\rm s-b}=\sum_{i,{\bm{k}}}g^i(a^{\dagger}_{\bm{k}}\otimes A^{i}_{\bm{k}}+a_{\bm k}\otimes A^{i{\dagger}}_{\bm{k}}+b^{\dagger}_{\bm{k}}\otimes B^{i}_{\bm{k}}+b_{{\bm{k}}}\otimes B^{i{\dagger}}_{\bm{k}}).
\end{equation}
Let us stress again that the strength of the coupling to each mode of the bath is represented by $g^i$. This, analogously to the energy of each mode $\epsilon^i$, is taken to be independent of the lattice site and of the type of bath $A$ or $B$, which is rather natural.

In terms of the operators $c_{\bm k}$ and $d_{\bm k}$, the Hamiltonian \eqref{Hintk} reads as
\begin{equation}
H_{\rm s-b}=\sum_{i,{\bm{k}}}g^i(c^{\dagger}_{\bm{k}}\otimes C^{i}_{\bm{k}}+c_{\bm k}\otimes C^{i{\dagger}}_{\bm{k}}+d^{\dagger}_{\bm{k}}\otimes D^{i}_{\bm{k}}+d_{{\bm{k}}}\otimes D^{i{\dagger}}_{\bm{k}}).
\end{equation}
where
\begin{equation}
\begin{pmatrix} C^i_{\bm{k}}\\ D^i_{\bm{k}} \end{pmatrix}=\frac{1}{\sqrt{1+|x_{\bm{k}}|^2}}\begin{pmatrix} x^{\ast}_{\bm{k}} & -1\\ 1 & x_{\bm{k}} \end{pmatrix}\begin{pmatrix} A^i_{\bm{k}}\\ B^i_{\bm{k}} \end{pmatrix}
\end{equation}
are new fermionic modes of the bath. Moreover note that
\begin{equation}
H_{\rm b}=\sum_{i,\bm{k}}\epsilon^i\left(C^{i\dagger}_{\bm{k}}C^i_{\bm{k}}+D^{i\dagger}_{\bm{k}}D^i_{\bm{k}}\right).
\end{equation}
Now, it is easy to write $H_{\rm s-b}$ in the interaction picture and apply the formula \eqref{weakCoupling}, which can be quite simplified by using that
\begin{eqnarray*}
{\rm Tr}_{\rm b}(C^{j\dagger}_{\bm{k'}}C^i_{\bm{k}}\rho_\beta)&=&{\rm Tr}_{\rm b}(D^{j\dagger}_{\bm{k'}}D^i_{\bm{k}}\rho_\beta)=\bar{n}_F(\epsilon^i)\delta_{i,j}\delta_{{\bm{k}},{\bm{k'}}},\\
{\rm Tr}_{\rm b}(C^{j}_{\bm{k'}}C^{i\dagger}_{\bm{k}}\rho_\beta)&=&{\rm Tr}_{\rm b}(D^{j}_{\bm{k'}}D^{i\dagger}_{\bm{k}}\rho_\beta)=[1-\bar{n}_F(\epsilon^i)]\delta_{i,j}\delta_{{\bm{k}},{\bm{k'}}},\\
{\rm Tr}_{\rm b}(C^{j\dagger}_{\bm{k'}}D^i_{\bm{k}}\rho_\beta)&=&{\rm Tr}_{\rm b}(D^{j\dagger}_{\bm{k'}}C^i_{\bm{k}}\rho_\beta)=0.
\end{eqnarray*}
Here, $\bar{n}_F(E):=\frac{1}{\text{e}^{\beta E}+1}$ is the mean number of particles of the Fermi-Dirac distribution, where we have taken the chemical potential $\mu$ to be at the origin of the energy. In the continuous limit for the baths degrees of freedom we have
\begin{equation}
\sum_i(g^i)^2f(\epsilon^i) \longrightarrow \int d\epsilon J(\epsilon)f(\epsilon),
\end{equation}
for any function $f(\epsilon)$, where $J(\omega)$ is the so-called spectral density of the bath. Thus, the Sokhotsky's identity
\begin{equation}\label{Sokhotsky}
\int^{\infty}_0d\tau {\rm e}^{{\rm i}\omega\tau}=\pi\delta(\omega)+{\rm i}{\rm PV}\left(\frac{1}{\omega}\right)
\end{equation}
allows us to simplify further the final expression, which after a bit long but straightforward computation reads
\begin{equation}\label{masterEqSM}
\begin{split}
&\frac{d\rho_{\rm s}(t)}{dt}=\sum_{\bm{k}}{\cal L}_{\bm{k}}[\rho_{\rm s}(t)] =\sum_{\bm{k}}\bigg(-{\rm i}[H_{\bm{k}},\rho_{\rm s}(t)] \\
&+\gamma(E_1^{\bm{k}})\bar{n}_F(E_1^{\bm{k}})\Big(c^{\dagger}_{\bm{k}}\rho_{\rm s}(t)c_{\bm{k}}-\frac{1}{2}\{c_{\bm{k}}c^{\dagger}_{\bm{k}},~\rho_{\rm s}(t)\}\Big)+\\
&+\gamma(E_1^{\bm{k}})[1-\bar{n}_F(E_1^{\bm{k}})]\Big(c_{\bm{k}}\rho_{\rm s}(t)c^{\dagger}_{\bm{k}}-\frac{1}{2}\{c^{\dagger}_{\bm{k}}c_{\bm{k}},\rho_{\rm s}(t)\}\Big) \\
&+\gamma(E_2^{\bm{k}})\bar{n}_F(E_2 ^{\bm{k}})\Big(d^{\dagger}_{\bm{k}}\rho_{\rm s}(t)d_{\bm{k}}-\frac{1}{2}\{d_{\bm{k}}d^{\dagger}_{\bm{k}},\rho_{\rm s}(t)\}\Big)+\\
&+\gamma(E_2^{\bm{k}})[1-\bar{n}_F(E_2^{\bm{k}})]\Big(d_{\bm{k}}\rho_{\rm s}(t)d^{\dagger}_{\bm{k}}-\frac{1}{2}\{d^{\dagger}_{\bm{k}}d_{\bm{k}},~\rho_{\rm s}(t)\}\Big)\bigg),
\end{split}
\end{equation}
in the Schr\"odinger picture, where $\gamma(\omega):=2\pi J(\omega)$. In addition, in this equation we have neglected the imaginary parts of Eq. \eqref{Sokhotsky} because they represent just a small shift of energies which does not affect to the dissipative process \cite{RivasPlato}.

\subsection{Cylindrical geometry}

In this case, we take Fourier transform along the direction $\bm{a}_2$ in \eqref{H}. Thus, the Haldane Hamiltonian reads as \eqref{HsCilinder}, the bath Hamiltonian becomes
\begin{equation}
H_{\rm b}=\sum_{k_2}\sum_{i,m}\epsilon^i\left(A^{i\dagger}_{(m,k_2)}A^i_{(m,k_2)}+B^{i\dagger}_{(m,k_2)}B^i_{(m,k_2)}\right),
\end{equation}
and the interaction Hamiltonian
\begin{equation}
\begin{split}
H_{\rm s-b} &=\sum_{k_2\in{\rm B.Z.}}\sum_{i,m}g^i\bigg(a^{\dagger}_{(m,k_2)}\otimes A^{i}_{(m,k_2)}+a_{(m,k_2)}\otimes A^{i{\dagger}}_{(m,k_2)}\\
&+b^{\dagger}_{(m,k_2)}\otimes B^{i}_{(m,k_2)}+b_{(m,k_2)}\otimes B^{i{\dagger}}_{(m,k_2)}\bigg).
\end{split}
\label{HintCylinder}
\end{equation}
where $m$ runs from 1 to the number of two-sites basic cells along the direction $\bm{a}_1$, $N_1$.

We may collect the operators $a_m$ and $b_m$ of each site in a new operator $c_m$ where $c_1:=a_1$, $c_2:=b_1$, $c_3:=a_2$, $c_4:=b_2$ and so on. The same can be done for the baths operators $A$ and $B$ with the notation $C_m$. Then, the interaction Hamiltonian \eqref{HintCylinder} is written as
\begin{equation}\label{HintCylinder2}
H_{\rm s-b}=\sum_{k_2\in{\rm B.Z.}}\sum_{i,m}g^i\Big(c^{\dagger}_{(m,k_2)}\otimes C^{i}_{(m,k_2)}+c_{(m,k_2)}\otimes C^{i{\dagger}}_{(m,k_2)}\Big).
\end{equation}
where now $m$ runs from 1 to $2N_1$.

The diagonal modes $f_{(m,k_2)}$ of $H(k_2)=\sum_{m}E_m(k_2)f^\dagger_{(m,k_2)}f_{(m,k_2)}$, where $E_m(k_2)$ is depicted in Fig. \ref{plotEdgeBands}, are related to $c_{(m,k_2)}$ by some unitary transformation
\begin{equation}
c_{(m,k_2)}=\sum_{\ell}w_{m,\ell}^{k_2}f_{(\ell,k_2)},\quad \text{with }  \sum_{\ell}w_{\ell,m}^{k_2\ast}w_{\ell,m'}^{k_2}=\delta_{m,m'}.
\end{equation}
By using this equation in \eqref{HintCylinder2} and after a bit of algebra we arrive at
\begin{equation}\label{HintCylinder3}
H_{\rm s-b}=\sum_{i,m,k_2}g^i\Big(f^{\dagger}_{(m,k_2)}\otimes F^{i}_{(m,k_2)}+f_{(m,k_2)}\otimes F^{i{\dagger}}_{(m,k_2)}\Big),
\end{equation}
where
\begin{equation}
F^{i}_{(m,k_2)}:=\sum_{\ell}w_{\ell,m}^{k_2\ast}C^{i}_{(\ell,k_2)},
\end{equation}
are new fermionic bath modes.

Following the same steps as for the toroidal geometry, it is not difficult to obtain the master equation of the cylindrical array,
\begin{equation}\label{masterEqCylinderSM}
\begin{split}
\frac{d\rho_{\rm s}(t)}{dt}=&\sum_{k_2\in {\rm B.Z.}}{\cal L}_{k_2}[\rho_{\rm s}(t)]=\sum_{k_2\in {\rm B.Z.} }\bigg(-{\rm i}[H(k_2),\rho_{\rm s}(t)]\\
+&\sum_{m}\Big(\gamma\big(E_m^{k_2}\big)\bar{n}_F\big(E_m^{k_2}\big) \mathcal{D}_{f^{\dagger}_{(m,k_2)}}[\rho_{\rm s}(t)]\\
+&\gamma\big(E_m^{k_2}\big)\big[1-\bar{n}_F\big(E_m^{k_2}\big)\big] \mathcal{D}_{f_{(m,k_2)}}[\rho_{\rm s}(t)]\Big)\bigg),
\end{split}
\end{equation}
where
\begin{equation}
\mathcal{D}_{K}[\rho_{\rm s}(t)]:=K\rho_{\rm s}(t)K^\dagger-\frac{1}{2}\{K^{\dagger}K,\rho_{\rm s}(t)\}.
\end{equation}

\section{Steady State for the Haldane Model}
\label{app_D}

\subsection{Toroidal Geometry}

The steady state of the previous band Liouvillian \eqref{masterEqSM} is the Gibbs state of the Hamiltonian $H_{\rm s}$,
\begin{equation}
\rho_{\beta}=\frac{\text{e}^{-\beta H_{\rm s}}}{Z},
\label{steady0}
\end{equation}
where $\beta=1/T$, with $T$ the temperature of the fermionic bath, and $Z={\rm Tr}(\text{e}^{-\beta H_{\rm s}})$ the partition function. To prove this, first note that $[H_{\rm s}, \rho_{\beta}]=0$ so we just need to care about the dissipator in \eqref{masterEqSM}.
In addition, since the number operators $c^{\dagger}_{\bm{k}}c_{\bm{k}}$ and $d^{\dagger}_{\bm{k}}d_{\bm{k}}$ commute with $c^{\dagger}_{\bm{k'}}$, $c_{\bm{k'}}$, $d^{\dagger}_{\bm{k'}}$, and $d_{\bm{k'}}$ if ${\bm{k}}\not={\bm{k'}}$, and we are left only with the part where the crystalline momenta in ${\cal L}_{\bm{k}}$ and $\rho^{\bm{k}}_{\rm ss}$ are the same, as the others trivially vanish. Since
\begin{equation}
{\rm e}^{\beta E}\bar{n}_F(E)=[1-\bar{n}_F(E)],
\end{equation}
and
\begin{equation}
\rho^{\bm{k}}_{\rm ss}=\bigg(\frac{{\rm e}^{-\beta E_1^{\bm{k}}c^{\dagger}_{\bm{k}}c_{\bm{k}}}}{1+\text{e}^{-\beta E_1^{\bm{k}}}}\bigg)\bigg(\frac{{\rm e}^{-\beta E_2^{\bm{k}}d^{\dagger}_{\bm{k}}d_{\bm{k}}}}{1+\text{e}^{-\beta E_2^{\bm{k}}}}\bigg),
\end{equation}
it is easy to prove that
\begin{equation}
\begin{split}
\mathcal{L}_{\bm k} (\rho^{\bm k}_{\rm ss})&=\bar{n}_F(E_1^{\bm k})\bigg(c_{\bm{k}}^{\dagger}\rho^{\bm k}_{\rm ss}c_{\bm{k}}-\frac{1}{2}\{c_{\bm{k}}c^{\dagger}_{\bm{k}},\rho^{\bm k}_{\rm ss}\}\bigg)\\
&+[1-\bar{n}_F(E_1^{\bm k})]\bigg(c_{\bm{k}}\rho^{\bm k}_{\rm ss}c^{\dagger}_{\bm{k}}-\frac{1}{2}\{c^{\dagger}_{\bm{k}}c_{\bm{k}},\rho^{\bm k}_{\rm ss}\}\bigg)\\
&+\bar{n}_F(E_2^{\bm k})\bigg(d^{\dagger}_{\bm{k}}\rho^{\bm k}_{\rm ss}d_{\bm{k}}-\frac{1}{2}\{d_{\bm{k}}d^{\dagger}_{\bm{k}},\rho^{\bm k}_{\rm ss}\}\bigg)\\
&+[1-\bar{n}_F(E_2^{\bm k})]\bigg(d_{\bm{k}}\rho^{\bm k}_{\rm ss}d^{\dagger}_{\bm{k}}-\frac{1}{2}\{d^{\dagger}_{\bm{k}}d_{\bm{k}},\rho^{\bm k}_{\rm ss}\}\bigg)=0.
\end{split}
\end{equation} 
Moreover, the state $\rho_\beta$ is the unique steady state of \eqref{masterEqSM} as the interaction Hamiltonian \eqref{HintSM} satisfies the irreducibility condition presented in \cite{Spohn2}.

In order to analyze some properties of $\rho_\beta$, let us write the density matrix $\rho^{\bm{k}}_{\rm ss}$ in the occupation basis of the two bands for a fixed momentum ${\bm{k}}$. The basis reads as $\ket{ij}_{\bm{k}}$, where $i=0,1$ and $j=0,1$ stand for the occupation of one-particle state in the lower and upper bands respectively. Thus, $\rho^{\bm{k}}_{\rm ss}$ is a $4\times4$ diagonal matrix,
\begin{equation}
\rho^{\bm k}_{\rm ss}={\rm diag}\begin{pmatrix} p^{\bm{k}}_{0000}, p^{\bm{k}}_{1010},  p^{\bm{k}}_{0101},p^{\bm{k}}_{1111} \end{pmatrix}
\end{equation}
with
\begin{equation}\label{steadyparts}
\begin{split}
p^{\bm{k}}_{0000}&:=\left[(1+\text{e}^{-\beta E_1^{\bm{k}}})(1+\text{e}^{-\beta E_2^{\bm{k}}})\right]^{-1},\\
p^{\bm{k}}_{1010}&:=p^{\bm{k}}_{0000} \text{e}^{-\beta E_1^{\bm{k}}},\\
p^{\bm{k}}_{0101}&:=p^{\bm{k}}_{0000} \text{e}^{-\beta E_2^{\bm{k}}},\\
p^{\bm{k}}_{1111}&:= p^{\bm{k}}_{0000} \text{e}^{-\beta E_1^{\bm{k}}}\text{e}^{-\beta E_2^{\bm{k}}}.
\end{split}
\end{equation}
\vspace*{1ex}

Since we set the origin of energy at $E_0=0$ and also took the chemical potential $\mu=0$ in between the two bands, $E_1^{\bm{k}}<0, E_2^{\bm{k}}>0$. Thus, at the low temperature limit, the state $\rho_{\rm ss}^{\bm{k}} \rightarrow\ket{10}_{\bm{k}}$ as $T\rightarrow0$ K. This means that at $T=0$ K the lower band is fully occupied and the upper band is completely empty, which is actually what one may expect. At the opposite limit, for $T\rightarrow\infty$, the system gets completely mixed, as the four possible states can be equally populated by the environment.

For the sake of clarity, we write the members of the occupation basis as
\begin{eqnarray*}
\ket{00}_{\bm{k}}&=&\ket{0}\ket{0},\\
\ket{10}_{\bm{k}}&=&\ket{u_{c,\bm{k}}}\ket{0},\\
\ket{01}_{\bm{k}}&=&\ket{0}\ket{u_{d,\bm{k}}},\\
\ket{11}_{\bm{k}}&=&\ket{u_{c,\bm{k}}}\ket{u_{d,\bm{k}}}.\\
\end{eqnarray*}
Then by using Eqs. \eqref{bandconnections} and \eqref{ArhoSimplified2}, we obtain (note that the $\partial_i\ket{0}=0$ as by definition the vacuum has no particles and so it does not depend on ${\bm{k}}$)
\begin{equation}\label{ArhoSimplified3}
A^\rho_i(\bm{k})=p_{1010}^{\bm{k}}A^{c}_i({\bm{k}})+p^{\bm{k}}_{0101}A^{d}_i({\bm{k}})+p^{\bm{k}}_{1111}(A^{c}_i({\bm{k}})+A^{d}_i({\bm{k}})).
\end{equation}
where the Berry connections for the lower $c$ and upper $d$ bands are provided by
\begin{equation}
A^{\alpha}_i({\bm{k}})={\rm i}\bra{u_{\alpha,\bm{k}}}\partial_i u_{\alpha,\bm{k}}\rangle, \quad \alpha = c, d.
\end{equation}
Thus, the expressions for $p^{\bm{k}}_{ijkl}$ given in Eq. \eqref{steadyparts} lead to
\begin{equation}
A^\rho_i(\bm{k})=\bar{n}_F(E_1^{\bm{k}})A^{c}_i({\bm{k}})+\bar{n}_F(E_2^{\bm{k}})A^{d}_i({\bm{k}}).
\end{equation}

\subsection{Cylindrical Geometry}

Because of the same reasons as with the toroidal geometry, the master equation on a cylindrical geometry \eqref{masterEqCylinderSM} has a unique steady state, which is the Gibbs state at the same temperature as the bath
\begin{equation}\label{steadyCylinderA}
\rho_{\beta}=\frac{\text{e}^{-\beta \sum_{k_2}H(k_2)}}{Z}=\bigotimes_{k_2} \frac{\text{e}^{-\beta H(k_2)}}{Z_{k_2}},
\end{equation}
with $Z_{k_2}=\traza \left[\text{e}^{-\beta H(k_2)}\right]$. Provided the system exhibits topological order, this state can be split as
\begin{equation}
\rho_\beta(k_2)=\frac{{\rm e}^{-\beta H(k_2)}}{Z_{k_2}}=\rho_\beta^{\rm L}(k_2)\otimes \rho_\beta^{\rm bulk}(k_2)\otimes \rho_\beta^{\rm R}(k_2),
\end{equation}
where
\begin{equation}
\rho_\beta^{\rm L,R}(k_2):=\frac{{\rm e}^{-\beta E_{L,R}(k_2)f^\dagger_{(L,R,k_2)}f_{(L,R,k_2)}}}{1+{\rm e}^{-\beta E_{L,R}(k_2)}},
\end{equation}
are Gibbs states involving just the gapless edge modes depicted in Fig. \ref{plotEdgeBands}.



%
%







\begin{thebibliography}{99}
\bibitem{footnote1} Physically, the ancillary space may represent the thermal bath. However, the dimension of the ancillary space has no physical consequences, it is just restricted to be larger or equal than the dimension of the system's Hilbert space in order to not lose generality.
\bibitem{comment} Note that in this formula we have restored the units $e$ (electron charge) and $h$ (Planck constant) as it is customary
in this context.
\end{thebibliography}
\end{document}